\documentclass[aps,showpacs,preprintnumbers,nofootinbibt,twocolumn]{revtex4-1}
\usepackage{amssymb}
\usepackage{graphicx}

\def\be{\begin{equation}}
\def\ee{\end{equation}}
\def\bea{\begin{eqnarray}}
\def\eea{\end{eqnarray}}

\begin{document}

\title{Cosmic strings in $f\left(R,L_m\right)$ gravity}
\author{Tiberiu Harko$^{1}$}
\email{t.harko@ucl.ac.uk}
\author{Matthew J. Lake$^{2,3}$}
\email{matthewj@nu.ac.th}
\affiliation{$^1$Department of Mathematics, University College London, Gower Street,
London, WC1E 6BT, United Kingdom}
\affiliation{$^2$ 
The Institute for Fundamental Study, ``The Tah Poe Academia Institute", \\
Naresuan University, Phitsanulok 65000, Thailand and \\
$^3$ Thailand Center of Excellence in Physics, Ministry of Education,
Bangkok 10400, Thailand }

\begin{abstract}
We consider Kasner-type static, cylindrically symmetric interior string
solutions in the $f\left(R,L_m\right)$ theory of modified gravity. The
physical properties of the string are described by an anisotropic
energy-momentum tensor satisfying the condition $T_t^t=T_z^z$; that is, the
energy density of the string along the $z$-axis is equal to minus the string
tension. As a first step in our study we obtain the gravitational field
equations in the $f\left(R,L_m\right)$ theory for a general static,
cylindrically symmetric metric, and then for a Kasner-type metric, in which
the metric tensor components have a power law dependence on the radial
coordinate $r$. String solutions in two particular modified gravity models
are investigated in detail. The first is the so-called ``exponential"
modified gravity, in which the gravitational action is proportional to the
exponential of the sum of the Ricci scalar and matter Lagrangian, and the
second is the ``self-consistent model", obtained by explicitly determining
the gravitational action from the field equations under the assumption of a power law dependent matter Lagrangian. In each case, the
thermodynamic parameters of the string, as well as the precise form of
the matter Lagrangian, are explicitly obtained. %
\end{abstract}

\maketitle



\section{Introduction} \label{Sect.I} 

The recently released Planck satellite data of the 2.7 $K$ Cosmic
Microwave Background (CMB) full sky survey \cite{Planckresults} have generally confirmed the standard $\Lambda $
Cold Dark Matter ($\Lambda $CDM) cosmological model.
Nonetheless, observations of high redshift type Ia supernovae have
convincingly shown that our universe is presently undergoing a period of
accelerated expansion, for which the most natural explanation would be the
presence of some form of dark energy, with an equation of state parameter $%
w=-1.018\pm 0.057$ for a flat universe \cite{Bet}. From a theoretical point
of view, the necessity of explaining dark energy, as well as the second
dominant component of the universe, dark matter \cite{Str}, raises the
fundamental question of whether the standard Einstein$-$Hilbert action ($S=\int%
{\left(R/2+L_m\right)\sqrt{-g}d^4x}$, where $R$ is the scalar curvature, and
$L_m$ is the matter Lagrangian density), in which matter is minimally
coupled to geometry, can give an appropriate quantitative description of the
universe on all scales, from the boundary of the solar system to the edge of
the horizon.

A gravitational theory with an explicit coupling between an arbitrary
function of the scalar curvature and the Lagrangian density of matter was
proposed in \cite{1}. The gravitational action of this theory is of the form
$S=\int{\left\{f_1(R)+\left[1+\lambda f_2(R)\right]L_m\right\}\sqrt{-g}d^4x}$%
, where $\lambda $ is a constant, and $f_1(R)$ and $f_2(R)$ are arbitrary
functions of the Ricci scalar. In this model an extra force acting on
massive test particles arises, and the motion is no longer geodesic.
Moreover, in this framework, one can also explain dark matter \cite{1}. The
initial linear geometry$-$matter coupling introduced in \cite{1} was
extended in \cite{2}, and a maximal extension of the Einstein$-$Hilbert action
with geometry-matter coupling, of the form $S=\int d^{4}x \sqrt{-g}%
f\left(R,L_m\right)$, was considered in \cite{fL}. Geometry$-$matter couplings
in the presence of scalar fields were discussed in \cite{fLS} and the
cosmological and astrophysical implications of the $f\left(R,L_m\right)$
type gravity theories were investigated in \cite{litfLm}. For a recent
review of the $f\left(R,L_m\right)$ model see \cite{Gal}.

Topological defects, including strings, are expected to have formed via the
Kibble mechanism \cite{Ki:76} during in phase transitions in the early
Universe \cite{Ki:80,Ki:82,topological_defects}. In particular, magnetic
monopoles are generic features of Grand Unified Theories (GUTs) \cite%
{Pr:84,Pr:87}, with expected mass scales in the region of $m \sim
m_{GUT}/\alpha_{GUT} \sim 10^{17}$ GeV, where $m_{GUT} \sim 10^{15}$ GeV is
the expected grand unification scale and $\alpha_{GUT}$ is the coupling
constant. Though not completely ruled out by existing data, the mass and
number densities of monopoles and domain walls, resulting from the
spontaneous breaking of spherical and parity symmetry, respectively, are
tightly constrained by observations \cite{Gu:81,topological_defects,
ParkerBound,Ad_etal,Ze:75, Ze:78,Ma:91,Na:91}.

For the time being, however, we limit ourselves to consideration of the most
viable species of topological defects which may exist as remnants from the
early universe; cosmic strings. In field theory models, strings form
whenever an axial symmetry is spontaneously broken. They are either
``infinitely" long (that is, spanning the horizon), or exist in the form of
loops. As shown by both analytical studies \cite{OneScaleAnalytic} and
numerical simulations \cite{OneScaleNumerical}, cosmic string networks
typically reach a scaling solution in which their contribution to the total
energy density of the universe becomes constant. This is due to string
self-intersection and the resulting ``chopping off" of loops from the long
string network, which then subsequently decay via gravitational radiation
\cite{StringGravRad} or gauge particle emission \cite{StringGaugeRad}.

Remarkably, this behavior also seems to occur in models motivated by string
theory, in which cosmic $F$-strings (fundamental, Nambu$-$Goto strings \cite%
{NambuGoto} of macroscopic size) and $D$-strings (one-dimensional $D$-branes
\cite{StringTheory}) or, more generally, bound states of $p$ $F$-strings and
$q$ $D$-strings called $(p,q)$-strings, exist in space-times with compact
extra dimensions \cite{FDStrings}. In such a scenario, one may expect the
inter-commuting probability of the string network to be significantly less
than unity, leading to reduced loop formation \cite{Thesis} but, to date,
predictions of scaling behavior based on analytic and numerical studies
appear to be robust \cite{ScalingFDStrings}.

Though there are a great many field-theoretic and string theory inspired
models of cosmic string formation, evolution and decay within the existing
literature, at present, almost all studies of the gravitational properties
of strings have been conducted in the context of general relativity or, at
least, in models for which nonstandard gravitational effects (such as those
caused by the presence of compact extra dimensions) may be neglected.
Notable exceptions include a handful of studies in $f(R)$ gravity \cite%
{f(R)_strings,f(R)_strings*, f(R)_stringsHD}, teleparallel theories \cite%
{Teleparallel_strings}, brane worlds \cite{Braneworld_strings}, Kaluza$-$Klein
models \cite{KK_strings}, Lovelock \cite{Lovelock_strings}, Gauss$-$Bonnet
\cite{GB_strings}, Born$-$Infeld \cite{BI_strings} and bimetric \cite%
{Bimetric_strings} gravity theories, and a few other, less mainstream,
alternatives \cite{Misc_strings}, while more comprehensive bodies of work
exists for strings in scalar-tensor theories \cite{ScalarTensor_strings},
including string theory-inspired dilation gravity \cite{Dilaton_strings},
and gravitational theories with torsion \cite{Torsion_strings}. The possible existence of Bose-Einstein condensate strings was investigated in \cite{BECstring}.

It is the
goal of the present paper to consider string-type solutions in the $%
f\left(R,L_m\right)$ gravity theory and, to the best of the author's
knowledge, this study constitutes the first analysis of such
solutions in a gravitational theory with arbitrary nonminimal
matter-geometry coupling.

Of the string-type solutions obtained in $(3+1)$-dimensional modified
gravity theories, those found in general $f(R)$ gravity \cite%
{f(R)_strings,f(R)_strings*,f(R)_strings**} bear closest resemblance to the solutions
considered in the present work. In \cite{f(R)_strings}, Azadi \emph{et al}
considered cylindrically symmetric solutions with constant Ricci curvature.
They found the unique solution for $R=0$, which is also recovered in Sects. %
\ref{SectV}-\ref{SectVI} as a specific case, corresponding to
cylindrically symmetric metrics obeying both sets of Kasner conditions \cite%
{Kasner,exact-sol}. In addition, they found two further families of vacuum
solutions for which $R = \mathrm{const. }\neq 0$, in which the Ricci scalar
was found to play the role of a cosmological constant term, giving rise to
an $f(R)$ gravity analogue of the Linet$-$Tian solution \cite{LinetTian}, for
cosmic strings in an Einstein gravity space-time with $\Lambda \neq 0$. In
\cite{f(R)_strings*}, Momeni and Gholizade showed that the solution
previously obtained in \cite{f(R)_strings} is, in fact, only one member of
the most generalized Tian family in general relativity and, as such, is
physically applicable to the exterior of a cosmic string.

However, the results derived in the present study seem, at first, to contrast with those
presented in \cite{f(R)_strings,f(R)_strings*}, since we determine the Ricci
scalar for general, Kasner-type space-times in $f(R,L_m)$ gravity to be of
the form $R = R_0(a,b,c)/r^2$, where $R_0(a,b,c)=0$ when both sets of Kasner
conditions are satisfied. Since the $f(R,L_m)$ theory naturally encompasses
all cases in which the gravitational Lagrangian is of the form $f(R) + L_m$,
it necessarily implies that cylindrically symmetric vacuum solutions in the $%
f(R)$ theory, for which $R = {\rm const.} \neq 0$, cannot be of Kasner form. By
the results presented in \cite{f(R)_strings*}, the physical implication of
this statement is that Linet$-$Tian type solutions in $f(R)$ gravity, in which the
cosmological constant term in the field equations is ``generated" by the constant Ricci scalar,
\emph{cannot} be described by the Kasner metric, and more general
cylindrically symmetric metrics must be considered in order to obtain them.
This contrasts with the case in general relativity, in which a ``Kasner-type" solution,
corresponding to $R = 4\Lambda = {\rm const.}$ exists  \cite{Spindel:79}, but is not of the standard Kasner form
(i.e obeying both sets of Kasner conditions, for which the unique vacuum solution
to Einstein's field equations with $\Lambda=0$ is recovered).

Some additional, nonvaccuum, solutions in $f(R)$ gravity were also found in \cite{f(R)_strings**},
using a self-consistent method in which the form of $f(R)$ was determined
from the general field equations using a cylindrically symmetric metric.
Though not generally of Kasner form, a more careful comparison of these solutions with the nonvacuum
solutions obtained in the present work warrants further study, but is beyond the cope of this paper.

For vacuum strings in the wire approximation \cite{Anderson}, or genuine $F$%
-strings governed by the Nambu$-$Goto action \cite{NambuGoto}, in which the
string world-sheet carries no additional currents or fluxes, the condition $%
\mathcal{T} = -\tilde{m}$, where $\tilde{m}$ is the inertial mass per unit
length and $\mathcal{T}$ is the tension, holds for a static, straight string
(commonly assumed to lie parallel to the $z$-axis). This condition arises
from general Lorentz invariance and, specifically, from boost invariance
along the string length. For vacuum strings of finite width and,
potentially, nonuniform mass densities in the radial direction, $r$ (such
as the Abelian-Higgs string \cite{no}), the equivalent condition is $%
T^{t}_{t} = T^{z}_{z}$, or $\rho(r) = -p_z(r)$, where $\rho(r)$ and $p_z(r)$
denote the mass density and thermodynamic pressure per unit volume within
the string core. Since, for a cylindrically symmetric distribution of
matter, these quantities depend only on the radial coordinate, integration
over $r$ yields the effective mass per unit length, $\tilde{m} = \int T^{tt}%
\sqrt{-g}dr$, and tension $\mathcal{T} = \int T^{zz}\sqrt{-g}dr$, of the
string-type field configuration, which thus also satisfy $\mathcal{T} =
-\tilde{m}$ \cite{VaVo:06}. For the sake of simplicity, we restrict our
attention to vacuum strings throughout the following analysis and impose the
condition $T^{t}_{t} = T^{z}_{z}$ on the matter sector.

The structure of this paper is then as follows. In Sect.~\ref{Sect.II}, we
give a brief outline of the $f(R,L_m)$ gravity theory, and we derive the
general field equations and conservation equations from the action. The form
of these equations is determined for a general static,
cylindrically symmetric source in Sect.~\ref{Sect.IIIA}. Specific Kasner
type solutions \cite{Kasner,exact-sol} of the field equations are presented
in Sect.~\ref{Sect.IIIB}. In Sect.~\ref{Sect.IV}, a specific choice is
made for the function $f$, and the exponential modified gravity theory, $%
f\left(R,L_m\right)=\Lambda \exp \left[(1/2\Lambda) R+(1/\Lambda)L_m\right]$%
, which reduces naturally to Einstein gravity in the low energy limit, is
chosen as a toy model. Solutions with self-consistent matter Lagrangians are
derived in Sect.~\ref{SectV}. That is, the self-consistent form of $%
f(R,L_m)$ is determined by imposing the condition $T^{t}_{t} = T^{z}_{z}$ on
the matter sector, which may be seen as a physical requirement corresponding
to the existence of vacuum/Nambu$-$Goto strings,
and by invoking the correspondence principle, which states that the standard
general relativistic limit of the modified gravity action (i.e. the
Einstein$-$Hilbert action), must exist for appropriate choices of the
arbitrary parameters and functions of the model. Specific, Kasner-type
solutions with self-consistent matter Lagrangians are investigated in
Sect.~\ref{SectVI}, for a simple choice of matter Lagrangian
dependence in $f$, $f(R,L_m) \propto L_m^q$, $q={\rm constant}\neq 1$, and two distinct string-type
solutions are found to exist. These correspond to two self-consistent sets
of constraints, which may be imposed on the free parameters of the Kasner
metric. Section~\ref{SectVII} contains a brief summary of the conclusions
and discussion of the main results of this analysis. In the present paper we use the natural system of units, with $8\pi \mathcal{G}=c_{light}=1$, where $c_{light}$ denotes the speed of light, and $\mathcal{G}$ the gravitational constant, respectively.

\section{$f\left(R,L_m\right)$ gravity} \label{Sect.II}

In this section, we first briefly summarize the $f\left(R,L_m\right)$ theory
of gravity and introduce the gravitational action, the field equations and
the conservation equations, respectively. The $f\left( R,L_{m}\right) $
theory is based on the assumption that the gravitational Lagrangian is given
by an arbitrary function $f$ of the Ricci scalar $R$ and of the matter
Lagrangian $L_{m}$ \cite{fL}. This theory represents a maximal extension of
the Hilbert$-$Einstein action, and the action takes the form \cite{fL}
\begin{equation}
S=\int f\left( R,L_{m}\right) \sqrt{-g}\;d^{4}x~,
\end{equation}%
where
$\sqrt{-g}$ is the square root of the determinant of the metric tensor. We
require that the function $f$ be analytic in both $R$ and $L_m$; that is, $f$
must possess a Taylor series expansion about any point $\left(R,L_m\right)$.
The energy-momentum tensor of the matter is defined by \cite{LaLi}
\begin{equation}
T_{\mu \nu }=-\frac{2}{\sqrt{-g}}\left[ \frac{\partial \left( \sqrt{-g}%
L_{m}\right) }{\partial g^{\mu \nu }}-\frac{\partial }{\partial x^{\lambda }}%
\frac{\partial \left( \sqrt{-g}L_{m}\right) }{\partial \left( \partial
g^{\mu \nu }/\partial x^{\lambda }\right) }\right] .
\end{equation}%
Thus, by assuming that the Lagrangian density of the matter, $L_{m}$,
depends only on the metric tensor $g_{\mu \nu }$, and not on its
derivatives, we obtain $T_{\mu \nu }=g_{\mu \nu }L_{m}-2\;\partial
L_{m}/\partial g^{\mu \nu }$.

By varying the action with respect to the metric we obtain the following
field equations for the $f\left( R,L_{m}\right) $ gravity theory \cite{fL}:
\begin{eqnarray}  \label{field2a}
&&\hspace{-0.3cm}f_{R}\left( R,L_{m}\right) R_{\mu \nu }+\left( g_{\mu \nu }\Box -\nabla
_{\mu }\nabla _{\nu }\right) f_{R}\left( R,L_{m}\right) 
\nonumber\\
&&\hspace{-0.3cm}-\frac{1}{2}\Bigg[f\left( R,L_{m}\right) -f_{L_{m}}\left( R,L_{m}\right)
L_{m}\Bigg]g_{\mu \nu } =
\frac{1}{2}f_{L_{m}}\left( R,L_{m}\right) T_{\mu \nu },\nonumber\\
\end{eqnarray}%
where we have denoted $f_{R}\left( R,L_{m}\right)=\partial f\left(
R,L_{m}\right)/\partial R$ and $f_{L_m}\left( R,L_{m}\right)=\partial
f\left( R,L_{m}\right)/\partial L_m$, respectively. Alternatively, Eq.~(\ref%
{field2a}) can be written as
\begin{eqnarray}
R_{\mu \nu } &=&\frac{1}{f_{R}\left( R,L_{m}\right) }\hat{P}_{\mu \nu
}f_{R}\left( R,L_{m}\right) +\Lambda_{eff} \left( R,L_{m}\right) g_{\mu \nu
} \nonumber\\
&&+ G_{eff}\left( R,L_{m}\right) T_{\mu \nu },  \label{field3}
\end{eqnarray}%
where we have denoted
\begin{equation}
{\hat{P}}_{\mu \nu }=\nabla _{\mu }\nabla _{\nu }-g_{\mu \nu }\Box ,
\end{equation}%
\begin{equation}  \label{Leff}
\Lambda_{eff} \left( R,L_{m}\right) =\frac{1}{2f_{R}\left( R,L_{m}\right) }%
\Bigg[f\left( R,L_{m}\right) -f_{L_{m}}\left( R,L_{m}\right) L_{m}\Bigg],
\end{equation}%
and
\begin{equation}  \label{Geff}
G_{eff}\left( R,L_{m}\right) =\frac{1}{2}\frac{f_{L_{m}}\left(
R,L_{m}\right) }{f_{R}\left( R,L_{m}\right) }.
\end{equation}%
The differential operators $\nabla _{\mu }\nabla _{\nu }$ and $\Box $ are
given by
\begin{equation}
\nabla _{\mu }\nabla _{\nu }F(x^{\lambda})=\frac{\partial ^{2}F}{\partial
x^{\mu }\partial x^{\nu }}-\Gamma _{\mu \nu }^{\alpha }\frac{\partial F}{%
\partial x^{\alpha }},
\end{equation}%
and
\begin{equation}
\Box F(x^{\lambda})=\frac{1}{\sqrt{-g}}\frac{\partial }{\partial x^{\mu }}%
\left( \sqrt{-g}g^{\mu \nu }\frac{\partial F}{\partial x^{\nu }}\right),
\end{equation}%
where $F\left(x^{\lambda}\right) $ is an arbitrary function of the
coordinates $x^{\lambda}$, and the $\Gamma _{\mu \nu }^{\alpha }$, defined as
\begin{equation}
\Gamma _{\alpha \mu \nu }=\frac{1}{2}\left( \frac{\partial g_{\alpha \mu }}{%
\partial x^{\nu }}+\frac{\partial g_{\alpha \nu }}{\partial x^{\mu }}-\frac{%
\partial g_{\mu \nu }}{\partial x^{\alpha }}\right) ,
\end{equation}%
are the Christoffel symbols associated to the metric.

With the choice $f\left( R,L_{m}\right) =R/2+L_{m}$ (the Einstein$-$Hilbert
Lagrangian), we recover the standard Einstein field equations of general
relativity, $R_{\mu \nu }-(1/2)g_{\mu \nu }R=T_{\mu \nu }$. By choosing $%
f\left( R,L_{m}\right) =f_{1}(R)+f_{2}(R)G\left( L_{m}\right) $, where $%
f_{1} $ and $f_{2}$ are arbitrary functions of the Ricci scalar, and $G$ is
a function of the matter Lagrangian density, we may reobtain the field
equations of the modified gravity with arbitrary curvature-matter coupling,
introduced in \cite{1}, for an appropriate choice of $G$.

The contraction of Eq.~(\ref{field2a}) gives the following relation between
the Ricci scalar $R$, the matter Lagrangian density $L_{m}$, and the trace
of the energy-momentum tensor, $T=T_{\mu }^{\mu }$;
\begin{eqnarray}
&&\hspace{-1cm}R+\frac{3}{f_{R}\left( R,L_{m}\right) }\Box f_{R}\left( R,L_{m}\right)
-4\Lambda \left( R,L_{m}\right) \nonumber\\
&&=G_{eff}\left( R,L_{m}\right) T.  \label{contr3a}
\end{eqnarray}

By eliminating the $\Box f_{R}\left( R,L_{m}\right)$ term using Eqs.~(\ref%
{field2a}) and (\ref{contr3a}), we obtain the gravitational field equations
in an alternative form as
\begin{eqnarray}
&&\hspace{-0.5cm}R_{\mu \nu }-\frac{1}{3}Rg_{\mu \nu }+\frac{1}{3}\Lambda \left(
R,L_{m}\right) g_{\mu \nu } =G_{eff}\left( R,L_{m}\right)\times \nonumber\\
&&\hspace{-0.5cm}\left( T_{\mu \nu }-\frac{1}{3}Tg_{\mu \nu }\right) + \frac{1}{f_{R}\left(
R,L_{m}\right) }\nabla _{\mu }\nabla _{\nu }f_{R}\left( R,L_{m}\right).
\end{eqnarray}

Taking the covariant divergence of Eq.~(\ref{field2a}), with the use of the
mathematical identity \cite{Koi},
\begin{eqnarray}  \label{koi}
\Box\nabla_{\nu}-\nabla_{\nu}\Box = R_{\mu\nu}\nabla^{\mu},
\end{eqnarray}
then gives the following equation for the divergence of the energy-momentum
tensor $T_{\mu \nu}$,
\begin{eqnarray}
\nabla ^{\mu }T_{\mu \nu } &=& \nabla ^{\mu }\ln \left[ f_{L_m}\left(R,L_m%
\right)\right] \left\{ L_{m}g_{\mu \nu }-T_{\mu \nu }\right\}
\nonumber\\
&&= 2\nabla ^{\mu }\ln \left[ f_{L_m}\left(R,L_m\right) \right] \frac{%
\partial L_{m}}{\partial g^{\mu \nu }}\,.  \label{noncons1}
\end{eqnarray}
The explicit derivation of Eq.~(\ref{noncons1}) is also presented in
Appendix \ref{app}. The requirement that the energy-momentum tensor for
matter be conserved, $\nabla ^{\mu }T_{\mu \nu }=0$, then yields an
effective functional relation between the matter Lagrangian density and the
total Lagrangian $f\left(R,L_m\right)$,
\begin{equation}
\nabla ^{\mu }\ln \left[ f_{L_m}\left(R,L_m\right) \right] \frac{\partial
L_{m}}{\partial g^{\mu \nu }}=0\,.
\end{equation}
Thus, once the matter Lagrangian density is known, by an appropriate choice
of $f\left( R,L_{m}\right) $, one can construct, at least in principle,
conservative gravity models with arbitrary curvature-matter coupling.

\section{Field equations for $f\left(R,L_m\right)$ modified gravity assuming
a static cylindrical symmetric source} \label{Sect.III} 

In the present section we consider the gravitational field equations
describing explicit string solutions in the $f\left(R,L_m\right)$ gravity
theory. We assume a cylindrically symmetric geometry in which the source
term is represented in the form of an anisotropic fluid with three distinct
pressures components along the $r$-, $\phi$- and $z$-axes (in cylindrical
polar coordinates, $(t,r,\phi,z)$). The simplest model for a cosmic string
is the so-called ``gauge string" \cite{KibbleH}, which is an idealized
cylindrical mass distribution with a finite radial extension having $T
_t^t=T_z^z$ as the only nonvanishing components of the energy-momentum
tensor. In the following we restrict our analysis to a similar
configuration, but with a more general pressure distribution, for which $%
T_r^r\neq 0$ and $T_{\phi }^{\phi}\neq 0$, respectively.

\subsection{String geometry and field equations in $f\left(R,L_m\right)$ gravity} \label{Sect.IIIA}

We assume a cylindrical symmetry metric, giving a line element of the form
\cite{clv}
\begin{equation}
ds^{2}=N^{2}(r)dt^{2}-d{r}^{2}-L^{2}(r)d{\phi }^{2}-K^{2}(r)dz^{2},
\label{lineelement}
\end{equation}%
where $N(r)$, $L(r)$ and $K(r)$ are arbitrary functions of the radial
coordinate $r$. The nonzero Christoffel symbols associated to the metric (%
\ref{lineelement}) are given by
\begin{eqnarray}
&&\Gamma _{10}^{0}=\frac{N^{\prime }(r)}{N(r)},\Gamma
_{00}^{1}=N(r)N^{\prime }(r),\Gamma _{12}^{2}=\frac{L^{\prime }(r)}{L(r)},
\nonumber \\
&&\Gamma _{22}^{1}=-L(r)L^{\prime }(r),\Gamma _{13}^{3}=\frac{K^{\prime }(r)%
}{K(r)},\Gamma _{33}^{1}=-K(r)K^{\prime }(r),  \nonumber \\
&&
\end{eqnarray}%
where a prime denotes the derivative with respect to $r$. The operator $\Box
$ is given by
\begin{equation}
\Box =-\frac{1}{N(r)L(r)K(r)}\frac{d}{dr}\left[ N(r)L(r)K(r)\frac{d}{dr}%
\right] ,
\end{equation}%
while for the covariant derivatives of an arbitrary function $f(r)$ we
obtain
\begin{eqnarray*}
&&\hspace{-1cm}\nabla _{t}\nabla _{t}f =-N(r)N^{\prime }(r)\frac{df}{dr},\nabla
_{r}\nabla _{r}f=\frac{d^{2}f}{dr^{2}},  \nonumber \\
&&\hspace{-1cm}\nabla _{\phi }\nabla _{\phi }f =L(r)L^{\prime }(r)\frac{df}{dr},\nabla
_{z}\nabla _{z}f=K(r)K^{\prime }(r)\frac{df}{dr}.
\end{eqnarray*}

With the line element (\ref{lineelement}), the components of the Ricci
tensor are \cite{clv}
\begin{eqnarray}  \label{Ricci}
&&R_{t}^{t}=\frac{(LKN^{\prime })^{\prime }}{NLK} ,R_{r}^{r}=\frac{N^{\prime
\prime }}{N}+\frac{L^{\prime \prime }}{L}+\frac{K^{\prime \prime }}{K},\nonumber\\
&&R_{\phi }^{\phi }=\frac{(NKL^{\prime })^{\prime }}{NLK} ,
R_{z}^{z}=\frac{%
(NLK^{\prime })^{\prime }}{NLK}.
\end{eqnarray}

The source term in the field equations is given by the energy-momentum
tensor with the following components \cite{clv}:
\begin{eqnarray}  \label{Tmunu}
&&T_{t}^{t} =\rho (r), T_{r}^{r} =-p_{r}(r),
T_{\phi }^{\phi } =-p_{\phi}(r),\nonumber\\
&&T_{z}^{z} =-p_{z}(r)=\rho (r).
\end{eqnarray}

Therefore the field equations describing cylindrically symmetric string-type
solutions in $f\left( R,L_{m}\right) $ gravity can be written as
\begin{eqnarray}  \label{eq1s}
&&\hspace{-0.5cm}\frac{(LKN^{\prime })^{\prime }}{NLK}=G_{eff}\left( R,L_{m}\right) \rho
+\Lambda \left( R,L_{m}\right) -\frac{1}{f_{R}\left( R,L_{m}\right) } \nonumber\\
&&\hspace{-0.5cm} \times\left[ \frac{1}{NLK}\frac{d}{dr}%
\left( NLK\frac{d}{dr}\right) +\frac{N^{\prime }}{N}\frac{d}{dr}\right]
f_{R}\left( R,L_{m}\right) ,
\end{eqnarray}%
\begin{eqnarray}  \label{eq2s}
&&\hspace{-0.5cm}\frac{N^{\prime \prime }}{N}+\frac{L^{\prime \prime }}{L}+\frac{K^{\prime
\prime }}{K}=-G_{eff}\left( R,L_{m}\right) p_{r}+\Lambda \left(
R,L_{m}\right) \nonumber\\
&&\hspace{-0.5cm}+ \frac{1}{f_{R}\left( R,L_{m}\right) }\left[ \frac{1}{NLK}\frac{d}{dr}%
\left( NLK\frac{d}{dr}\right) -\frac{d^{2}}{dr^{2}}\right] f_{R}\left(
R,L_{m}\right) ,\nonumber\\
\end{eqnarray}%
\begin{eqnarray}  \label{eq3s}
&&\frac{(NKL^{\prime })^{\prime }}{NLK}=-G_{eff}\left( R,L_{m}\right)
p_{\phi }+\Lambda \left( R,L_{m}\right) \nonumber\\
&&+ \frac{1}{f_{R}\left( R,L_{m}\right) }\left[ \frac{1}{NLK}\frac{d}{dr}%
\left( NLK\frac{d}{dr}\right) -\frac{L^{\prime }}{L}\frac{d}{dr}\right] \nonumber\\
&& \times f_{R}\left( R,L_{m}\right) ,
\end{eqnarray}%
\begin{eqnarray}  \label{eq4s}
&&\frac{(NLK^{\prime })^{\prime }}{NLK}=G_{eff}\left( R,L_{m}\right) \rho
+\Lambda \left( R,L_{m}\right) \nonumber\\
&&+ \frac{1}{f_{R}\left( R,L_{m}\right) }\left[ \frac{1}{NLK}\frac{d}{dr}%
\left( NLK\frac{d}{dr}\right) -\frac{K^{\prime }}{K}\frac{d}{dr}\right] \nonumber\\
&& \times f_{R}\left( R,L_{m}\right) .
\end{eqnarray}

Generally, the regularity of the geometry on the symmetry axis is imposed
via the initial conditions,
\begin{equation}
L(0)=0,L^{\prime }(0)=1,N(0)=1,N^{\prime }(0)=0.  \label{initcond}
\end{equation}

We define the radius $R_s$ of the string as the radius for which the radial
pressure $p_r(r)$ vanishes, $p_r\left(R_s\right)=0$. Additionally, we may
also impose the conditions $p_{\phi}\left(R_s\right) = 0$, and $%
\int_{r=0}^{R_s}p_{r}(r) dr = 0$, and $\int_{r=0}^{R_s}p_{\phi}(r) dr = 0$,
respectively. The physical reasoning behind these conditions is that, if
they did not hold, the string core would either be expanding or
contracting. However, for some string models these conditions may not hold,
thus allowing the possibility of a nonzero surface energy density and
tangential pressure.

Next we consider the relations that follows from the ``conservation" of the
energy-momentum tensor, given by Eq.~(\ref{noncons1}). Since all the
components of the energy-momentum tensor and of the function $f\left(
R,L_{m}\right) $ are independent of the coordinates $\left( t,\phi ,z\right)
$, we have
\begin{equation}
\nabla _{\mu }T_{t}^{\mu }=0,\nabla _{\mu }T_{\phi }^{\mu }=0,\nabla _{\mu
}T_{z}^{\mu }=0.  \label{divzero}
\end{equation}
Furthermore, the divergence of the $T_{\nu }^{\mu }$ can be obtained in a
general form as \cite{LaLi}
\begin{equation}
\nabla _{\mu }T_{\nu }^{\mu }=\frac{1}{\sqrt{-g}}\frac{\partial }{\partial
x^{\mu }}\left( \sqrt{-g}T_{\nu }^{\mu }\right) -\frac{1}{2}\frac{\partial
g_{\alpha \beta }}{\partial x^{\nu }}T^{\alpha \beta }.  \label{divem}
\end{equation}

It follows that Eq.~(\ref{divzero}) is identically satisfied, with the only
potentially nonzero component of the divergence being given by
\begin{equation}
\nabla _{\mu }T_{r}^{\mu }=\frac{1}{\sqrt{-g}}\frac{d}{dr}\left( \sqrt{-g}%
T_{r}^{r}\right) -\frac{1}{2}\frac{\partial g_{\alpha \beta }}{\partial r}%
T^{\alpha \beta }.  \label{divem1}
\end{equation}

With the use of Eq. (\ref{divem1}), the ``conservation equation" for the
cosmic string in the $f\left( R,L_{m}\right)$ gravity theory takes
the form
\begin{eqnarray}  \label{divemf}
&&-\frac{dp_{r}}{dr}-\left( \rho +p_{r}\right) \frac{N^{\prime }}{N}+\left(
p_{\phi }-p_{r}\right) \frac{L^{\prime }}{L}+\left( p_{z}-p_{r}\right) \frac{%
K^{\prime }}{K}=\nonumber\\
&&\frac{d}{dr}\ln f_{L_m}\left( R,L_{m}\right) \left( L_{m}+p_{r}\right) .
\end{eqnarray}
and, since $p_z=-\rho $, this is equivalent to
\begin{eqnarray}  \label{43}
&&-\frac{dp_{r}}{dr}-\left( \rho +p_{r}\right) \left(\frac{N^{\prime }}{N}+%
\frac{K^{\prime }}{K}\right)+\left( p_{\phi }-p_{r}\right) \frac{L^{\prime }%
}{L}=\nonumber\\
&&\frac{d}{dr}\ln f_{L_m}\left( R,L_{m}\right) \left( L_{m}+p_{r}\right) .
\end{eqnarray}

To obtain a description of the physical properties of the string we
introduce the Tolman mass per unit length $M(r)$ of this system, defined as \cite{def}
\begin{eqnarray}
M (r)&=&2\pi \int_{0}^{r }{\left( \rho +p_{r}+p_{\phi }+p_{z}\right)
N^{2}Ldr}=\nonumber\\
&&2\pi \int_{0}^{r }{\left( p_{r}+p_{\phi }\right) N^{2}Ldr},
\end{eqnarray}%
along with the parameter
\begin{eqnarray}
W(r) &=&-2\pi \int_{0}^{r }{\left( \rho -p_{r}+p_{\phi }-p_{z}\right)
N^{2}Ldr}=\nonumber\\
&&-2\pi \int_{0}^{\infty }{\left( 2\rho -p_{r}+p_{\phi }\right) N^{2}Ldr},
\end{eqnarray}%
which can be related to the angular deficit of the space-time via the
``angular" Einstein equation \cite{def}. 
With the use of the field equations Eqs. (\ref{eq1s})$-$(\ref{eq4s}) we obtain
\begin{widetext}
\begin{eqnarray}
M(r) &=&2\pi \int_{0}^{r }\frac{1}{G_{eff}\left( R,L_{m}\right) }\left\{
2\Lambda _{eff}\left( R,L_{m}\right) +\frac{1}{f_{R}\left( R,L_{m}\right) }\left[
\frac{2}{NLK}\frac{d}{dr}\left( NLK\frac{d}{dr}\right) -\frac{d^{2}}{dr^{2}}-%
\frac{L^{\prime }}{L}\frac{d}{dr}\right] f_{R}\left( R,L_{m}\right) \right\}
N^{2}Ldr- \nonumber\\
&&2\pi \int_{0}^{\infty }\frac{1}{G_{eff}\left( R,L_{m}\right) }\left[ \frac{%
N^{\prime \prime }}{N}+\frac{L^{\prime \prime }}{L}+\frac{K^{\prime \prime }%
}{K}+\frac{(NKL^{\prime })^{\prime }}{NLK}\right] N^{2}Ldr,
\end{eqnarray}
and
\begin{eqnarray}
W(r) &=&-2\pi \int_{0}^{r }\frac{1}{G_{eff}\left( R,L_{m}\right) }\left[
\frac{(LKN^{\prime })^{\prime }}{NLK}+\frac{N^{\prime \prime }}{N}+\frac{%
L^{\prime \prime }}{L}+\frac{K^{\prime \prime }}{K}-\frac{(NKL^{\prime
})^{\prime }}{NLK}+\frac{(NLK^{\prime })^{\prime }}{NLK}\right] N^{2}Ldr- \nonumber\\
&&2\pi \int_{0}^{\infty }\frac{1}{G_{eff}\left( R,L_{m}\right) }\left\{2\Lambda _{eff}\left(R,L_m\right)-\frac{1}{f_{R}\left(
R,L_{m}\right) }\left[ \frac{d^{2}}{dr^{2}}+\left( \frac{N^{\prime }}{N}-%
\frac{L^{\prime }}{L}+\frac{K^{\prime }}{K}\right) \frac{d}{dr}\right]f_R\left(R,L_m\right)\right\} N^{2}Ldr,\nonumber\\
\end{eqnarray}
\end{widetext}
respectively. In the following we denote the total value of the Tolman mass and the value of the angular deficit parameter on the string surface by $M=M\left(R_s\right)$ and $W=W\left(R_s\right)$, respectively.

\subsection{The string gravitational field equations for the Kasner metric} \label{Sect.IIIB} 

We now investigate the possibility of the existence of Kasner-type solutions
for static, cylindrically symmetric strings in $f\left(R,L_m\right)$
gravity. In standard general relativity, by neglecting the effect of the
cosmological constant, the unique, static, cylindrically symmetric vacuum
solution of the Einstein field equations is given by the Kasner metric \cite%
{Kasner, exact-sol},
\begin{eqnarray}  \label{Kasner1}
ds^{2} = (kr)^{2a}dt^{2} - dr^{2} - \beta^{2} (kr)^{2(b-1)}r^{2}d{\phi}^2-
(kr)^{2c}dz^{2} ,\nonumber\\
\end{eqnarray}
where $k$ sets the length scale and $\beta$ is a constant, related to the
deficit angle of the conical space-time. The Kasner metric is characterized
by two free parameters which, for the unique vacuum solution, satisfy the Kasner conditions \cite%
{Kasner,exact-sol},
\begin{equation}
a + b + c=a^2 + b^2 + c^2=1.  \label{Kasner2}
\end{equation}

The derivation of the Kasner metric in standard general relativity is
presented in Appendix \ref{app1}. The physical interpretation of the free
parameters $a$, $b$, $c$ of the Kasner metric is of fundamental importance
in the context of the description of cosmic strings, as is the relation
between these parameters and the internal properties of the matter
distribution. The simplest model for a cosmic string, the gauge string
\cite{KibbleH}, belongs to a very simple class of Kasner-type solutions,
with $a=c=0, b=1$, which is locally flat. However, as one can see from Eq.~(%
\ref{Kasner1}), in the gauge string model, there is an important global
non-trivial effect, namely, that the string geometry is that of a cone. From
a geometrical point of view the parameter $\beta$ in the Kasner solution is
directly related to the conic angular deficit \cite{Marder2,Bonnor}, which,
in turn, is also related to the mass distribution of the source \cite%
{FuGa:88,GaLa:89,LaGa:89,Ra:90,BoCh:90}.

In the case of a gauge string (in general relativity), there is an explicit relation between the
angular deficit $\Delta \phi =2\pi (1-\beta )$ and the inertial mass (per
unit length) $\tilde{m}$, given by $\Delta \phi =8\pi G\tilde{m}$. This
relation was first obtained by using a linear approximation for the case of
an infinitesimally thin source \cite{Vil1}. The same relation was obtained
by solving the full nonlinear Einstein equations around a uniform source ($%
T_{t}^{t}=\mathrm{constant}$) with a finite radius \cite{Gott,Hiscock}, and
in the case of a nonuniform source \cite{Linet1}. Since the space-time
around a gauge string is locally flat, the angular deficit is the only
geometrical (and physical) evidence of its existence, thus allowing the
possibility of observationally testing the presence of cosmic strings. The
next step in the study of gravitating cosmic strings requires the analysis
of the more realistic, Abelian-Higgs type models. In this context the
analysis of the full coupled field equations for the gravitational field and
matter fields was performed in \cite{Garfinkle1}.

Hence, for a Kasner-type metric, the gravitational field equations for the
cylindrically symmetric static string in $f(R,L_m)$ gravity (\ref{eq1s})-(\ref{eq4s})
take the form
\begin{eqnarray}  \label{ff1}
&&\frac{\left( a+b+c-1\right) a}{r^{2}}=G_{eff}\left( R,L_{m}\right) \rho
+\Lambda_{eff} \left( R,L_{m}\right) \nonumber\\
&&- \frac{1}{f_{R}\left( R,L_{m}\right) }\left[ \frac{d^{2}}{dr^{2}}+\frac{%
2a+b+c}{r}\frac{d}{dr}\right] f_{R}\left( R,L_{m}\right) ,
\end{eqnarray}%
\newline
\begin{eqnarray}  \label{ff2}
&&\frac{a^{2}+b^{2}+c^{2}-\left( a+b+c\right) }{r^{2}}=-G_{eff}\left(
R,L_{m}\right) p_{r}\nonumber\\
&&+ \Lambda_{eff} \left( R,L_{m}\right) +\frac{a+b+c}{r}\frac{d}{dr}\ln
f_{R}\left( R,L_{m}\right) ,
\end{eqnarray}%
\begin{eqnarray}  \label{ff3}
&&\frac{\left( a+b+c-1\right) b}{r^{2}}=-G_{eff}\left( R,L_{m}\right)
p_{\phi }+\Lambda \left( R,L_{m}\right) \nonumber\\
&&+ \frac{1}{f_{R}\left( R,L_{m}\right) }\Bigg[\frac{d^{2}}{dr^{2}}+\frac{a+c}{%
r}\frac{d}{dr}\Bigg]f_{R}\left( R,L_{m}\right) ,
\end{eqnarray}%
\begin{eqnarray}  \label{ff4}
&&\frac{\left( a+b+c-1\right) c}{r^{2}}=G_{eff}\left( R,L_{m}\right) \rho
+\Lambda_{eff} \left( R,L_{m}\right) \nonumber\\
&&+ \frac{1}{f_{R}\left( R,L_{m}\right) }\left[ \frac{d^{2}}{dr^{2}}+\frac{a+b%
}{r}\frac{d}{dr}\right] f_{R}\left(R,L_{m}\right) .
\end{eqnarray}
The conservation equation~(\ref{43}) becomes
\begin{eqnarray}  \label{cons2}
&&-\frac{dp_{r}}{dr}-\left( \rho +p_{r}\right) \frac{a+c}{r}+\left( p_{\phi
}-p_{r}\right) \frac{b}{r} \nonumber\\
&&= \frac{d}{dr}\ln f_{L_m}\left( R,L_{m}\right) \left( L_{m}+p_{r}\right).
\end{eqnarray}
and the Ricci scalar is given by $R=R_{0}/r^{2}$, where
\begin{equation}
R_{0}=a^{2}+b^{2}+c^{2}+\left( a+b+c\right) \left( a+b+c-2\right).
\label{ricci}
\end{equation}

The Tolman mass per unit length $M(r)$ and the $W(r)$ parameter, which controls the
angular deficit of the string in the Kasner metric, are obtained as
\begin{widetext}
\begin{eqnarray}
M(r) &=&2\pi \beta k\int_{0}^{r }\frac{\left( 1-b\right) (a+b+c)+b-\left(
a^{2}+b^{2}+c^{2}\right) }{G_{eff}\left( R,L_{m}\right) }\left( kr\right)
^{2a+b-2}dr+  \nonumber \\
&&\frac{2\pi \beta }{k}\int_{0}^{\infty }\frac{\left( kr\right) ^{2a+b}}{%
G_{eff}\left( R,L_{m}\right) }\left\{ 2\Lambda _{eff}\left( R,L_{m}\right) +\frac{1%
}{f_{R}\left( R,L_{m}\right) }\left[ \frac{d^{2}}{dr^{2}}+\frac{2a+b+2c}{r}%
\frac{d}{dr}\right] f_{R}\left( R,L_{m}\right) \right\} dr
\end{eqnarray}%
and
\begin{eqnarray}
W(r) &=&-2\pi k\beta \int_{0}^{r }\frac{1}{G_{eff}\left( R,L_{m}\right) }%
\left[ \left( a+b+c-1\right) \left( a-b+c-1\right) +a^{2}+b^{2}+c^{2}-1%
\right] \left( kr\right) ^{2a+b-2}dr- \nonumber\\
&&\frac{2\pi \beta }{k}\int_{0}^{\infty }\frac{\left( kr\right) ^{2a+b}}{%
G_{eff}\left( R,L_{m}\right) }\left\{ 2\Lambda _{eff}\left( R,L_{m}\right) -%
\frac{1}{f_{R}\left( R,L_{m}\right) }\left[ \frac{d^{2}}{dr^{2}}+\frac{a-b+c%
}{r}\frac{d}{dr}\right] f_{R}\left( R,L_{m}\right) \right\} dr,
\end{eqnarray}
\end{widetext}
respectively.

\section{Kasner-type string solutions with a given form of the Lagrangian
density: Exponential $f\left(R,L_m\right)$ gravity} \label{Sect.IV} 

As a first example of string-type cylindrically symmetric solutions in $%
f\left(R,L_m\right)$ gravity we consider a Lagrangian density of the form
\cite{fL}
\begin{equation}  \label{exp}
f\left(R,L_m\right)=\Lambda \exp \left(\frac{1}{2\Lambda }R+\frac{1}{\Lambda
}L_m\right),
\end{equation}
\newline
where $\Lambda >0$ is an arbitrary constant. In the limit $\left( 1/2\Lambda
\right) R+\left( 1/\Lambda \right) L_{m}\ll 1$, we obtain
\begin{equation}
f\left( R,L_{m}\right) \approx \Lambda +\frac{R}{2}+L_{m}+...
\end{equation}
\newline
That is, we recover the full Einstein$-$Hilbert gravitational Lagrangian with
a cosmological constant.

With this choice of Lagrangian density the gravitational field equations
take the form
\begin{widetext}
\begin{eqnarray}  \label{expp1}
&&R_{\mu \nu }=\left( \Lambda -L_{m}\right) g_{\mu \nu }+T_{\mu \nu }-
\frac{1}{\Lambda }\left[ \left( \frac{1}{2}\nabla _{\mu
}\nabla ^{\mu} R+\nabla _{\mu} \nabla ^{\mu } L_{m}\right) g_{\mu
\nu}\right.-
\left.\left( \frac{1}{2}\nabla _{\mu }\nabla _{\nu }R+\nabla _{\mu }\nabla
_{\nu}L_{m}\right) \right]  \nonumber\\
&&- \frac{1}{\Lambda ^{2}}\left[ \left( \frac{1}{2}\nabla ^{\lambda }R+\nabla
^{\lambda }L_{m}\right) \left( \frac{1}{2}\nabla _{\lambda }R+\nabla
_{\lambda }L_{m}\right) g_{\mu \nu }\right.-
\left.\left( \frac{1}{2}\nabla _{\mu }R+\nabla _{\mu }L_{m}\right) \left(
\frac{1}{2}\nabla _{\nu}R+\nabla _{\nu }L_{m}\right) \right] .
\end{eqnarray}
\end{widetext}

In the case of weak gravitational fields and of small particle velocities,
the exponential curvature-matter coupling induces an extra acceleration of
massive test particles, which is proportional to the gradients of both the
Ricci scalar and the matter Lagrangian, $\vec{a}=-\left( 1/\Lambda \right) %
\left[ \nabla \left( R/2\right) +\nabla L_{m}\right] $. In the case of
pressureless dust, the extra force is proportional to the gradient of the
matter density $\rho $ only.

As one can see from Eqs.~(\ref{expp1}), the gravitational field equations of
the exponential $f\left(R,L_m\right)$ gravity model contain a new background term, proportional to the metric
tensor $g_{\mu \nu }$, which depends on both the constant term $\Lambda $,
and on the physical parameters of the matter. Thus, through the
geometry$-$matter coupling, the exponential model introduces an effective,
time dependent ``cosmological constant''.

From Eqs.~(\ref{Leff}) and (\ref{Geff}), we have
\begin{equation}
\Lambda _{eff} \left(R,L_m\right)=\Lambda -L_m(r),
G_{eff}\left(R,L_m\right)=1.
\end{equation}

Then, for a Kasner-type metric, the field equations (\ref{ff1})$-$(\ref{ff4})
describing the static, cylindrically symmetric, string in exponential $%
f\left( R,L_{m}\right) $ gravity take the form
\begin{widetext}
\begin{eqnarray}\label{exp1}
\frac{\left( a+b+c-1\right) a}{r^{2}} &=&\rho (r)+\Lambda -L_{m}(r)-\frac{%
2R_{0}-\Lambda r^{2}(2a+b+c)}{\Lambda ^{2}r^{3}}L_{m}^{\prime }(r)+\frac{%
R_{0}\left[ \Lambda r^{2}(2a+b+c-3)-R_{0}\right] }{\Lambda ^{2}r^{6}}-
\frac{L_{m}^{\prime \prime }(r)}{\Lambda }-\frac{L_{m}^{\prime 2}}{\Lambda
^{2}}, \nonumber\\
\end{eqnarray}%
\bea\label{exp2}
\frac{a^{2}+b^{2}+c^{2}-\left( a+b+c\right) }{r^{2}}=-p_{r}(r)+\Lambda
-L_{m}(r)+\frac{a+b+c}{\Lambda r^{4}}\left[ r^{3}L_{m}^{\prime }(r)-R_{0}%
\right] ,
\eea
\bea\label{exp3}
\frac{\left( a+b+c-1\right) b}{r^{2}}=-p_{\phi }(r)+\Lambda -L_{m}(r)+\frac{%
\left[ \Lambda r^{2}(a+c)-2R_{0}\right] }{\Lambda ^{2}r^{3}}L_{m}^{\prime
}(r)+\frac{L_{m}^{\prime 2}}{\Lambda ^{2}}+\frac{R_{0}\left[ R_{0}-\Lambda
r^{2}(a+c-3)\right] }{\Lambda ^{2}r^{6}}+\frac{L_{m}^{\prime \prime }(r)}{%
\Lambda },\nonumber\\
\eea
\bea\label{exp4}
\frac{\left( a+b+c-1\right) c}{r^{2}}=\rho(r) +\Lambda -L_{m}(r)+\frac{\Lambda
r^{2}(a+b)-2R_{0}}{\Lambda ^{2}r^{3}}L_{m}^{\prime }(r)+\frac{R_{0}\left[
R_{0}-\Lambda r^{2}(a+b-3)\right] }{\Lambda ^{2}r^{6}}+\frac{L_{m}^{\prime
\prime }(r)}{\Lambda }+\frac{L_{m}^{\prime 2}}{\Lambda ^{2}},
\eea
\end{widetext}
where we have also used the explicit form of the Ricci scalar for the Kasner
metric, given by Eq. (\ref{ricci}). Equations~(\ref{exp1}) and (\ref{exp4})
immediately give the following differential equation, which is satisfied by
the matter Lagrangian;
\begin{widetext}
\begin{equation}
\frac{L_{m}^{\prime \prime }(r)}{\Lambda }+\frac{L_{m}^{\prime 2}}{\Lambda
^{2}}-\frac{4R_{0}-\Lambda r^{2}(3a+2b+c)}{2\Lambda ^{2}r^{3}}L_{m}^{\prime
}(r)-\frac{R_{0}\left[ \Lambda r^{2}(3a+2b+c-6)-2R_{0}\right] }{2\Lambda
^{2}r^{6}}+\frac{\left( a+b+c-1\right) \left( c-a\right) }{2r^{2}}=0.
\label{expfin}
\end{equation}
\end{widetext}

By introducing a new dependent variable $l_{m}(r)$, defined via $%
L_{m}^{\prime }(r)=\Lambda l_{m}^{\prime }(r)/l_{m}(r)$, $L_{m}(r)=\Lambda
\ln l_{m}(r)$, Eq.~(\ref{expfin}) takes the form of a linear differential
equation for $l_{m}(r)$, given by
\begin{widetext}
\begin{equation}\label{lm}
l_{m}^{\prime \prime }(r)-\frac{4R_{0}-\Lambda r^{2}(3a+2b+c)}{2\Lambda r^{3}%
}l_{m}^{\prime }(r)-\left\{ \frac{R_{0}\left[ \Lambda r^{2}(3a+2b+c-6)-2R_{0}%
\right] }{2\Lambda ^{2}r^{6}}-\frac{\left( a+b+c-1\right) \left( c-a\right)
}{2r^{2}}\right\} l_{m}(r)=0.
\end{equation}
\end{widetext}

By introducing a new independent variable $\xi =1/r$, we have $l_{m}^{\prime
}=\left( dl_{m}/d\xi \right) \left( d\xi /dr\right) =-\xi ^{2}dl_{m}/d\xi $,
and $l_{m}^{\prime \prime }=2\xi ^{3}\left( dl_{m}/d\xi \right) +\xi
^{4}\left( d^{2}l_{m}/d\xi ^{2}\right) $, respectively. Therefore Eq.~(\ref%
{lm}) takes the form
\begin{eqnarray}
&&\frac{d^{2}l_{m}\left( \xi \right) }{d\xi ^{2}}-\frac{\left( \beta
-2\right) \xi ^{2}-s}{2\xi }\frac{dl_{m}\left( \xi \right) }{d\xi }-\nonumber\\
&&\left( \mu -\frac{\beta }{2}\xi ^{2}-\frac{m}{\xi ^{2}}\right) l_{m}\left(
\xi \right) =0,  \label{imp}
\end{eqnarray}%
where we have denoted $\beta =4R_{0}/\Lambda $, $s=3a+2b+c$, $\mu =\beta (s-6)/8$, and $%
m=(a+b+c-1)(c-a)/2$. Equation~(\ref{imp}) has the general solution 
\begin{widetext}
\begin{equation}\label{51}
l_{m}(\xi )=l_{m}^{0}e^{m_{1}\xi ^{2}}\xi ^{m_{2}}\left[ c_{1}U\left(
u_{1},u_{2},-\frac{1}{4}\sqrt{(\beta -12)\beta +4}\xi ^{2}\right)
+c_{2}L_{-u_{1}}^{u_{2}-1}\left( -\frac{1}{4}\sqrt{(\beta -12)\beta +4}\xi
^{2}\right) \right] ,
\end{equation}%
\end{widetext}where $U\left( a,b,z\right) $ is the confluent hypergeometric
function, defined as $U\left( a,b,z\right) =\left[ 1/\Gamma (a)\right]
\int_{0}^{\infty }e^{-zt}t^{a-1}\left( 1+t\right) ^{b-a-1}dt$, $L_{n}^{a}(x)$
is the generalized Laguerre polynomial, satisfying the differential equation
$xy^{\prime \prime }+(a+1-x)y^{\prime }+ny=0$, and where we have denoted
\begin{equation}
l_{m}^{0}=2^{\frac{1}{8}\left( \sqrt{(s-2)^{2}-16m}+4\right) },
\end{equation}%
\begin{equation}
m_{1}=\frac{1}{8}\left( \beta +\sqrt{(\beta -12)\beta +4}-2\right) ,
\end{equation}%
\begin{equation}
m_{2}=\frac{1}{4}\left( \sqrt{(s-2)^{2}-16m}-s+2\right) ,
\end{equation}%
\begin{widetext}
\begin{equation}
u_{1}=\frac{(s+2)\beta -2(s+4\mu +2)+\sqrt{(s-2)^{2}-16m}\sqrt{(\beta
-12)\beta +4}+4\sqrt{(\beta -12)\beta +4}}{8\sqrt{(\beta -12)\beta +4}},
\end{equation}%
\end{widetext}and
\begin{equation}
u_{2}=\frac{1}{4}\left( \sqrt{(s-2)^{2}-16m}+4\right).
\end{equation}%
The parameters $c_{1}$ and $c_{2}$ are arbitrary integration constants.
Hence, the thermodynamical variables of the Kasner string in the
exponential type $f\left( R,L_{m}\right) $ gravity can be expressed in terms
of the function $l_{m}(r)$ as
\begin{eqnarray}  \label{57}
\rho \left( r\right) &=&\Lambda \left[ \ln l_{m}(r)-1\right] +\frac{a+c}{2r}%
\frac{l_{m}^{\prime }(r)}{l_{m}(r)}\nonumber\\
&&+ \frac{\left( a+b+c-1\right) (3a-c)}{2r^{2}}-\frac{\left( a+c\right) R_{0}}{%
2\Lambda r^{4}},
\end{eqnarray}
\begin{eqnarray}  \label{58}
p_{r}(r)&=&\Lambda \left[ 1-\ln l_{m}(r)\right] +\frac{\left( a+b+c\right)
-\left( a^{2}+b^{2}+c^{2}\right) }{r^{2}}\nonumber\\
&&+ \frac{a+b+c}{\Lambda r^{4}}\left[ \Lambda r^{3}\frac{l_{m}^{\prime }(r)}{%
l_{m}(r)}-R_{0}\right] ,
\end{eqnarray}
\begin{eqnarray}  \label{59}
p_{\phi }\left( r\right) &=&\Lambda \left[ 1-\ln l_{m}(r)\right] +\frac{%
c-a-2b}{2r}\frac{l_{m}^{\prime }(r)}{l_{m}(r)}\nonumber\\
&&-\frac{\left( a+b+c-1\right) \left( c-2b-a\right) }{r^{2}}+\frac{\left(
a+b-2c\right) R_{0}}{2\Lambda r^{4}}.\nonumber\\
\end{eqnarray}

In the limit of small $\xi $, corresponding to large $r$, Eq.~(\ref{51}) can
be approximated as $l_m(\xi)\propto \xi ^{m_2}$, or, equivalently, $%
l_m(r)\propto r^{-m_2}$. Thus it follows that, at large radial distances, $%
L_m(r)\approx -m_2\Lambda \ln r$ and $l^{\prime }_m(r)/l_m(r)\approx -m_2/r$%
. By using these approximations for $L_m(r)$ and $l_m(r)$ in Eqs.~(\ref{57}%
)$-$(\ref{59}), we can easily determine the behavior of the energy density and
thermodynamic pressures of the Kasner string in the exponential $%
f\left(R,L_m\right)$ gravity theory.

In the same order of approximation we obtain for the gravitational
Lagrangian density and its derivative with respect to $R$ the expressions
\begin{eqnarray}
&&f\left(R,L_m\right)\approx \Lambda r^{-m_2}\exp \left(\frac{R_0}{2\Lambda
r^2}\right),\nonumber\\
&&f_R\left(R,L_m\right)\approx \frac{1}{2} r^{-m_2}\exp \left(\frac{R_0}{%
2\Lambda r^2}\right).
\end{eqnarray}

By using the results above we can now estimate the mass per unit length and
angular deficit parameter of the string in the exponential model. 
We assume that the string core extends between $r=0$ and $r=R_{s}$, and thus
we avoid the singularity at the upper integration limits in the integrals.
The string radius can be estimated from the equation $p_{r}\left(
R_{s}\right) =0$, which gives
\begin{eqnarray}  \label{radn}
&&\Lambda \left[ 1+m_{2}\ln R_{s}\right] +\frac{\left( a+b+c\right) -\left(
a^{2}+b^{2}+c^{2}\right) }{R_{s}^{2}}-\nonumber\\
&&\frac{a+b+c}{\Lambda R_{s}^{4}}\left[ \Lambda m_{2}R_{s}^{2}+R_{0}\right]
=0.
\end{eqnarray}

By neglecting the term $\Lambda \left[ 1+m_{2}\ln R_{s}\right] $ in Eq. (\ref%
{radn}), we obtain for the string radius the second-order algebraic equation
\bea
&&\hspace{-0.5cm}\left[ \left( a+b+c\right) -\left( a^{2}+b^{2}+c^{2}\right) \right]
R_{s}^{2}+m_{2}\left( a+b+c\right) R_{s}^{2}+\nonumber\\
&&\hspace{-0.5cm}(a+b+c)\frac{R_{0}}{\Lambda }=0.
\eea

If the Kasner conditions hold at least approximately, so that $a+b+c\approx
1$ and $a^{2}+b^{2}+c^{2}\approx 1$, the string radius can be approximated
as $R_{s}\approx \sqrt{-R_{0}/m_{2}\Lambda }$.

Hence, for the mass and the angular deficit parameter, we obtain
\begin{widetext}
\begin{eqnarray}\label{69}
M(r) &=&2\pi \beta \chi (a,b,c)(kr)^{2a+b-1}+\frac{4\pi \beta }{k^{2}}\frac{%
(kr)^{2a+b+1}}{2a+b+1}\Lambda -  2\beta k^{4}(kR_{s})^{2a+b-5}\Bigg[\frac{R_{0}R_{s}^{2}(2a+b+c-2m_{2}-3)}{%
\Lambda (2a+b-3)} \nonumber\\
&&+ \frac{m_{2}R_{s}^{4}(2a+b+c-m_{2}-1)}{2a+b-1}-\frac{2\Lambda m_{2}R_{s}^{6}%
}{(2a+b+1)^{2}}+ \frac{2\Lambda m_{2}R_{s}^{6}\log (R_{s})}{2a+b+1}-\frac{R_{0}^{2}}{%
\Lambda ^{2}(2a+b-5)}\Bigg]
\end{eqnarray}
\end{widetext}
and %
\begin{widetext}
\begin{eqnarray}
W(r)&=&-2\pi \beta \theta (a,b,c)(kr)^{2a+b-1}-\frac{4\pi \beta }{k^{2}}\frac{%
(kr)^{2a+b+1}}{2a+b+1}\Lambda +  2\pi \beta k^{4}(kR_{s})^{2a+b-5}\Bigg[ \frac{%
R_{0}R_{s}^{2}(-a+b-c+2m_{2}+3)}{\Lambda (2a+b-3)}  \nonumber \\
&&+ \frac{m_{2}R_{s}^{4}(-a+b-c+m_{2}+1)}{2a+b-1}+\frac{2\Lambda m_{2}R_{s}^{6}%
}{(2a+b+1)^{2}}-  \frac{2\Lambda m_{2}R_{s}^{6}\log (R_{s})}{2a+b+1}+\frac{R_{0}^{2}}{%
\Lambda ^{2}(2a+b-5)}\Bigg],
\end{eqnarray}
\end{widetext}
where
\be
\chi (a,b,c)=\left( 1-b\right) (a+b+c)+b-\frac{\left(
a^{2}+b^{2}+c^{2}\right)}{(2a+b-1)},
\ee
and
\bea\label{72}
&&\hspace{-0.85cm}\theta (a,b,c)=\nonumber\\
&&\hspace{-0.85cm}\frac{\left[ \left(
a+b+c-1\right) \left( a-b+c-1\right) +a^{2}+b^{2}+c^{2}-1\right]}{(2a+b-1)}.
\eea
In order to avoid any divergence at the origin, the parameters
$a$ and $b$ must satisfy the condition $2a+b>6$.

By assuming that the parameter $\Lambda $ in the exponential model is the cosmological constant, we may assume for it a numerical value of the order of $\Lambda \approx 10^{-56}\;{\rm cm ^{-2}}$. Then we can estimate the string radius as $R_s\approx \left(-R_0/m_2\right)^{1/2}\times 10^{28}$ cm. This relation imposes the constraint $R_0/m_2<0$. However, this value of the radius is obtained in the quasi-Kasner approximation, which implies that $R_0$ may have very small values, $R_0=\epsilon <<1$, while $m_2$ still can have values of the order of unity. Therefore the radius of the string can be written as $R_s\approx \sqrt{\epsilon}\times 10^{28}$ cm. To obtain physically reasonable values of the radius,  very small deviations from the exact Kasner regime are required. For example, with $\epsilon =10^{-40}$, we can obtain more realistic string radius values of the order of $R_s\approx 10^8$ cm.  By taking into account the approximate expression of $R_s$ in Eq.~(\ref{69}), we can estimate the mass per unit length of the string as
\bea\label{M1}
M&\approx &2\beta \left(\frac{c_{light}^2}{\mathcal{G}}\right)\left(kR_s\right)^{2a+b-1}\nonumber\\
&&\approx  2.7\times 10^{28}\times \beta \times \left(kR_s\right)^{2a+b-1}\;{\rm  g/cm}.
\eea
Besides the parameters $a$ and $b$, the mass of the string depends on the other two parameters $\beta $ and $k$ of the Kasner metric. For $k\approx 1/R_s$, and $\beta =10^{-4}$, we obtain for the mass per unit length of the string $M\approx 10^{24}$ g/cm. Therefore ultralong cosmic strings, with lengths of the order of $1\;{\rm  kpc}=3\times 10^{21}$ cm, could reach masses of order $M\approx 1.5\times 10^{12}M_{\odot}$ in this model. Of course these estimates are strongly dependent on the values of $\beta $, $k$ and the parameters of the Kasner metric, whose exact values can be obtained only by fitting with astrophysical data.

 In the first-order approximation the angular deficit parameter of the string can be written in a form very similar to the Tolman mass equation,
 \be
 W\approx -2\pi \beta \theta (a,b,c) \left(\frac {c_{light}^2}{\mathcal{G}}\right)\left(kR_s\right)^{2a+b-1}\;{\rm g/cm}.
 \ee
The basic difference between the numerical values of $M$ and $W$ is determined by the functional form of $\theta (a,b,c)$, which, as one can see from  Eq.~(\ref{72}), in the quasi-Kasner limit takes very small values, being exactly equal to zero in the standard Kasner case. So the presence of a nonzero negative angular deficit is determined by the deviation of the string interior metric from the exact Kasner form.

 \subsection{Observational implications} \label{ObsImp}

Gravitational lensing is one of the important physical effects that could, at least in principle, discriminate between cosmic strings and other matter distributions. In the standard string scenario a vacuum string does not change
the curvature of the space-time, but only its topology \cite{Gott}. Therefore light rays are not deflected by a cosmic
string. However, because of the specific conical structure of the space-time metric,
if two light rays pass on the different sides of the string, they
may converge later at the same point of observation \cite{Bozza}. If a cosmic
string lies between us and a distant source, we see two images
of the source separated by an angle $\delta \theta =8\pi \mu _s \sin \alpha D_{LS}/D_{OS}$, where $\mu _s$ is the string linear density, $\alpha $ is the angle between the string and the observer - source direction, and  $D_{LS}$ and
$D_{OS}$ indicate the distance  between the lens
and the source, and the observer and source, respectively. Hence in the case of the standard conical string, the two images formed due to the string presence are identical to the original source, without  presenting any amplification or distortion \cite{Bozza}. By contrast, the gravitational lensing by gas filaments shows a very different pattern, with one or three elongated images \cite{Bozza}.

The gravitational lensing patterns of the $f\left(R,L_m\right)$ cosmic strings are also fundamentally different from the lensing of the conical strings, being closer to the lensing patterns produced by long gas filaments. The metric outside the $f\left(R,L_m\right)$ cosmic string is of Kasner-type, and therefore the propagation of light is strongly affected by the space-time geometry. By assuming, in the first approximation,  that the interior density profile of the $f\left(R,L_m\right)$ cosmic string can be approximated by a Gaussian distribution with width $\sigma _0$, for the lensing angle of a light ray passing at a distance $D$ from
the string central axis, consisting of a massive cylindrical matter distribution, we obtain the expression $\alpha \approx \left(\mathcal{G}M/c_{light}^2\right){\rm Erf}\left(D/\sigma _0\sqrt {2}\right)$, where ${\rm Erf} (z)$ is the error function \cite{Bozza}. This equation is valid under the assumption that the length of the string is much larger than $D$. Hence, with the use of Eq.~(\ref{M1}), we obtain for the light deflection angle the expression
\be
\alpha \approx 2\beta \left(kR_s\right)^{2a+b-1}{\rm Erf}\left(\frac{D}{\sigma _0\sqrt {2}}\right).
\ee
As one can see from the above equation, the lensing angle therefore depends on the intrinsic parameters $k$ and $\beta $ of the metric, as well as on the Kasner coefficients $a$ and $b$. There is also a strong dependence on the string radius $R_s$, as well as on the matter density profile inside the string core. Due to the presence of a larger number of parameters in the deflection angle formula, cosmic strings in exponential $f\left(R,L_m\right)$ gravity could potentially provide a better fit to the observational data as compared to the standard string model. Also these massive string solutions predict a much richer lensing behavior, as compared to the simple conical string model, leading to the possibility of a much wider range of deflection angles.

Different aspects of lensing by cosmic strings have been analyzed in detail in the literature \cite{Saz, Lens}. A potential lensing effect, which may provide some evidence for the existence of cosmic strings, comes from the Capodimonte$-$Sternberg lens candidate (CSL-1) \cite{Saz}. However, detailed comparisons of observations with existing string lensing models have proved, ``beyond any doubts" \cite{Saz}, that the CSL-1 images are simply a rather peculiar pair of interacting elliptical galaxies. Nonetheless, a future improvement in the precision of the observations could lead to the potential identification of $f\left(R,L_m\right)$ strings, in the sense that lensing from such strings may fit CSL-1 data for appropriate choices of the model parameters.



\section{String solutions in $f\left( R,L_{m}\right) $ gravity with
self-consistent matter Lagrangian}

\label{SectV} 

In the present section we consider some explicit solutions of the field
equations (\ref{eq1s})--(\ref{eq4s}), without imposing \textit{a priori} the
functional form of $f\left(R,L_m\right)$.

From Eqs.~(\ref{ff1}) and (\ref{ff4}) it immediately follows that the
function $f_{R}$ satisfies the second-order linear differential equation
\begin{equation}  \label{diffeq}
\frac{d^{2}}{dr^{2}}f_{R}+\frac{3a+2b+c}{r}\frac{d}{dr}f_{R}=\frac{\left(
a+b+c-1\right) \left( c-a\right) }{r^{2}}f_{R},
\end{equation}%
which has the general solution
\begin{equation}
f_{R}\left( r\right) =C_{1}^{\prime \prime }r^{\alpha _{1}}+C_{2}^{\prime
\prime }r^{\alpha _{2}},  \label{eqR}
\end{equation}%
where $C_{1}^{\prime \prime }$ and $C_{2}^{\prime \prime }$ are arbitrary
constants of integration and $\alpha _{1}$ and $\alpha _{2}$ are solutions
of the algebraic equation
\begin{equation}
\alpha ^{2}+\left( 3a+2b+c-1\right) \alpha -\left( a+b+c-1\right) \left(
c-a\right) =0,
\end{equation}
given by
\begin{eqnarray}  \label{alpha}
\alpha _{1,2} &=&\frac{1-3a-2b-c}{2} \nonumber\\
&&\pm \frac{\sqrt{\left( 3a+2b+c-1\right) ^{2}+4\left( a+b+c-1\right) (c-a)}}{2}.
\nonumber \\
\end{eqnarray}

Using $R=R_{0}/r^{2}$, where $R_0$ is given by Eq.~(\ref{ricci}), we can
rewrite Eq.~(\ref{eqR}) as a function of $R$ as
\begin{equation}
f_{R}\left( R,L_{m}\right) =C_{1}^{\prime }R^{-\alpha _{1}/2}+C_{2}^{\prime
}R^{-\alpha _{2}/2},
\end{equation}
where $C_{1}^{\prime }=R_{0}^{\alpha _{1}/2}C_{1}^{\prime \prime }$ and $%
C_{2}^{\prime }=R_{0}^{\alpha _{2}/2}C_{2}^{\prime \prime }$ giving,
immediately,
\begin{equation}  \label{f}
f\left( R,L_{m}\right) =C_{1}R^{n_{1}}+C_{2}R^{n_{2}}+g\left( L_{m}\right) ,
\end{equation}%
where $n_1=1-\alpha _1/2$, $n_2=1-\alpha _2/2$, $C_{1}=C_{1}^{\prime
}/\left( 1-\alpha _{1}/2\right) $, $C_{2}=C_{2}^{\prime }/\left( 1-\alpha
_{2}/2\right) $, and $g\left( L_{m}\right) $ is an arbitrary integration
function of the matter Lagrangian $L_{m}$. This encompasses the solution
obtained in \cite{f(R)_strings} (which corresponds to the condition $%
f(R=0)=0 $, giving $f(R)$ as a superposition of powers $R^{k}$, $k \geq 1$%
), when both sets of Kasner conditions, Eq. (\ref{Kasner2}), are satisfied and $%
g(L_m) = 0$. However, it is clearly more general and holds for $R \neq 0$, $%
g(L_m) \neq 0$.

The components of the string energy-momentum tensor can then be obtained as
\begin{eqnarray}  \label{p1}
\rho &=&\frac{1}{G_{eff}\left( R,L_{m}\right) }\Bigg[ \frac{\left(
a+b+c-1\right) c}{r^{2}}-\Lambda _{eff}\left( R,L_{m}\right) \nonumber\\
 &&- \frac{a+b}{r}\frac{d}{dr}\ln f_{R}\left( R,L_{m}\right) \Bigg] ,
\end{eqnarray}
\begin{eqnarray}  \label{p2}
p_{r}&=&\frac{1}{G_{eff}\left( R,L_{m}\right) }\Bigg[ \frac{\left(
a+b+c\right) -\left( a^{2}+b^{2}+c^{2}\right) }{r^{2}} \nonumber\\
 &&+ \Lambda _{eff}\left( R,L_{m}\right) +\frac{a+b+c}{r}\frac{d}{dr}\ln
f_{R}\left( R,L_{m}\right) \Bigg] ,
\end{eqnarray}
\begin{eqnarray}  \label{p3}
p_{\phi }&=&\frac{1}{G_{eff}\left( R,L_{m}\right) }\Bigg[ \frac{\left(
a+b+c-1\right) \left( c-a-b\right) }{r^{2}}  \nonumber\\
&&+ \Lambda _{eff}\left( R,L_{m}\right) -\frac{2\left( a+b\right) }{r}\frac{d}{%
dr}\ln f_{R}\left( R,L_{m}\right) \Bigg] ,
\end{eqnarray}
\begin{equation}  \label{p4}
p_{z}=-\rho .
\end{equation}

By requiring that the generalized gravitational Lagrangian density 
must have the standard general relativistic limit, that is,
by requiring it to reduce to the standard Einstein$-$Hilbert Lagrangian for an
appropriate choice of the free parameters of the model, we can impose the
conditions
\begin{equation}
n_{1}=1,C_1=\frac{1}{2},
\end{equation}
which, according to Eq.~(\ref{f}), require $\alpha _1=0$. This criterion is
satisfied if the coefficients $a$, $b$, $c$ satisfy the condition
\begin{equation}  \label{condexp}
(a+b+c-1)(c-a)=0.
\end{equation}
Then, for $\alpha _2$ we obtain $\alpha _2=1-3a-2b-c$, giving the expression
\begin{equation}
n_2=n=\frac{\left(1+3a+2b+c\right)}{2}
\end{equation}
for $n_2=1-\alpha _2/2$, where, for simplicity, we rename $n_2$ as $n$ from
now on.

Therefore the self-consistent gravitational Lagrangian for a string in $%
f\left( R,L_{m}\right) $ gravity takes the form
\begin{equation}\label{Lm}
f\left( R,L_{m}\right) =\frac{R}{2}+\frac{C_{2}}{n}R^{n}+g\left(
L_{m}\right) ,
\end{equation}%
which reduces to the Einstein$-$Hilbert Lagrangian when $C_{2}=0$ and $g\left(
L_{m}\right) \rightarrow L_{m}$.

\subsection{The energy conservation equation} \label{SectVA} 

In order to completely solve the problem of the determination of the
generalized gravitational action for a string-like configuration, one must
determine the dependence of the function $f\left( R,L_{m}\right) $ on the
matter Lagrangian $L_{m}$. This can be achieved by substituting the
expressions of the energy-momentum tensor, given by Eqs.~(\ref{p1})$-$(\ref{p4}%
), into the energy conservation equation, Eq.~(\ref{cons2}). The resulting
equation completely determines the functional form of $f$, as well as the
dependence of the matter Lagrangian on the coordinate $r$. Taking into
account the conditions given by Eq.~(\ref{condexp}), we have to consider two
cases independently. \newline

\subsubsection{The case $a+b+c=1$, $c\neq a$} \label{SectVB} 

After substituting the expressions for the string energy density and
pressures into the conservation equation, Eq.~(\ref{cons2}), and by taking
into account the conditions $a+b+c=1$, $c\neq a$, we obtain the following
differential equation, giving the dependence of the function $%
g\left(L_m(r)\right)$ on the matter Lagrangian and on the radial coordinate $%
r$,
\begin{widetext}
\begin{eqnarray}\label{consa}
&&R_{0}\left\{ -2\left[ b^{2}+b(c-1)+(c-1)c\right] +r^{3}(-L_{m}(r))L_{m}^{%
\prime }(r)g^{\prime \prime }(L_{m}(r))+R_{0}\right\} \nonumber\\
&&- 2C_{2}r^{2}\left(
\frac{R_{0}}{r^{2}}\right) ^{n}\left[
b^{2}(4n-2)+2b(c-1)(5n-4)+c^{2}(4n-2)+c(4-6n)+4(n-1)n-R_{0}\right] =0.
\end{eqnarray}%
\end{widetext}

\subsubsection{The case $a+b+c\neq 1$, $c=a$} \label{SectVC} 

The second case in which the condition given by Eq.~(\ref{condexp}) is
satisfied is when the coefficients $a$, $b$, $c$ in the Kasner metric
satisfy the conditions $a+b+c\neq 1$ and $c=a$. For these values of the
Kasner coefficients the energy conservation equation Eq.~(\ref{cons2}) takes
the form
\begin{widetext}
\bea\label{consb}
&&R_{0}\left[ -6a^{2}-4a(b-1)-2(b-1)b+r^{3}(-L_{m}(r))L_{m}^{\prime
}(r)g^{\prime \prime }(L_{m}(r))+R_{0}\right] \nonumber\\
&&- 2C_{2}r^{2}\left( \frac{R_{0}%
}{r^{2}}\right) ^{n}\left[ 4n^{2}(2a+b)-2(2b+3)n(2a+b)+6(a+b)^{2}-R_{0}%
\right] =0.
\eea
\end{widetext}

\section{Kasner-type string solutions in $f\left(R,L_m\right)$ gravity with
self-consistent gravitational Lagrangian} \label{SectVI} 

In the present section we present some explicit, exact, string-type
solutions of the gravitational field equations in $f\left(R,L_m\right)$
gravity, for which the source term satisfies the condition $T_t^t=T_z^z$ and
the metric is assumed to be of the Kasner form. The solutions are obtained
by imposing specific conditions on the coefficients $a$, $b$, $c$, or by
assuming a specific functional form for the integration function $%
g\left(L_m(r)\right)$.

\subsection{Solutions satisfying both Kasner conditions} \label{SectVIA} 

We consider first string-type solutions that satisfy both the Kasner
conditions given in Eqs.~(\ref{Kasner2}). In this case $R_t^t=R_r^r=R_{%
\phi}^{\phi}=R_z^z=0$, and $R=0$, respectively. The conservation equation
Eq.~(\ref{consa}) reduces to the condition $g^{\prime \prime
}\left(L_m\right)=0$, giving $g\left(L_m\right)=\mathrm{constant}+L_m$.
Then, by taking into account the fact that $f_R\left(R,L_m\right)=0$, from
the gravitational field equations (\ref{field2a}), we obtain $T_{\mu \nu}=0$%
, implying that $L_m=0$. We therefore obtain again, in the framework of $%
f\left(R,L_m\right)$ gravity theory, the Kasner vacuum solution of standard
general relativity.

\subsection{Kasner-type string solutions with $g\left(L_m\right)=g_0L_m^q(r)$} \label{SectVIB} 

As a second example of a string solution in $f\left(R,L_m\right)$ gravity we
consider the simple case in which $g\left(L_m\right)$ has a power law dependence on $L_m$, so that $g\left(L_m\right)=g_0L_m^q(r)$, where $g_0$ and $q$ are constants.

With this choice of the matter term the conservation equation~(\ref%
{consa}) and Eq. (\ref{consb}) give the $r$-dependence of the matter
Lagrangian, allowing us to obtain, explicitly, the full solution of the
gravitational field equations for the Kasner string. In the following we
consider the cases $a+b+c=1$, $c\neq a$ and $a+b+c\neq 1$, $c=a$ separately.

\subsubsection{The case $a+b+c=1$, $c\neq a$} \label{SectVIB1} 

If the values of the constants $a$, $b$, $c$ satisfy the constraints $%
a+b+c=1 $, $c\neq a$, then from Eq.~(\ref{consa}), for $g\left(L_m%
\right)=g_0L_m^q(r) $, we obtain first the differential equation satisfied by the matter Lagrangian as
\begin{widetext}
\begin{eqnarray}
&&-2C_{2}r^{2}L_{m}(r)\left( \frac{R_{0}}{r^{2}}\right) ^{n}\left[
b^{2}(4n-2)+2b(c-1)(5n-4)+c^{2}(4n-2)+c(4-6n)+4(n-1)n-R_{0}\right]
\nonumber\\
&&+ R_{0}L_{m}(r)\left( R_{0}-2\left( b^{2}+b(c-1)+(c-1)c\right) \right)
-g_{0}(q-1)qr^{3}R_{0}L_{m}^{q}(r)L_{m}^{\prime }(r)=0,
\end{eqnarray}%
\end{widetext}
with the general solution given by
\begin{widetext}
\begin{equation}
L_{m}(r)=2^{-1/q}\left\{ \frac{r^{-2(n+1)}\left( 2C_{1}Ar^{2}-nR_{0}r^{2n}%
\left[ -2\left( b^{2}+b(c-1)+c^{2}\right) -2c_{1}g_{0}(q-1)qr^{2}+2c+R_{0}%
\right] \right) }{g_{0}n(q-1)R_{0}}\right\} ^{1/q},
\end{equation}
\end{widetext}
where $c_1$ is an arbitrary constant of integration, and  we have denoted
\bea
A&=&R_{0}^{n}\Bigg[ b^{2}(4n-2)+2b(c-1)(5n-4)\nonumber\\
&&+c^{2}(4n-2)+c(4-6n)+4(n-1)n-R_{0}\Bigg].
\eea

Hence we obtain the gravitational Lagrangian density explicitly as
\begin{equation}
f\left(R,L_m\right)=\frac{R}{2}+\frac{2}{2+2a+b}C_2R^{(2+2a+b)/2}+g_0L_m^q.
\end{equation}
Imposing $2a+b=2$ or, equivalently, $c=a-1$, and setting $C_2 = 1/\Lambda$,
we obtain the model $f\left(R,L_m\right) = \frac{1}{2}\left(R +
R^2/\Lambda^2\right)+g_0L_m^q$, whose $f(R)$ version with matter minimally coupled to
geometry is commonly studied  \cite{f(R)_strings}. Hence, static, string-type solutions \emph{do} exist in the
$f\left(R,L_m\right)$ cosmology, but their self-consistent description requires $R \propto r^{-2}
\neq {\rm constant}$ within the string core.

\subsection{The case $a+b+c\neq 1$, $c= a$} \label{SectVIC} 

If the parameters of the Kasner metric satisfy the conditions $a+b+c\neq 1$,
$c=a$, then for $g\left( L_{m}\right) =g_0L_{m}^q$, from the conservation Eq.~(%
\ref{consb}), it follows that  the matter Lagrangian satisfies the differential equation
\begin{widetext}
\begin{eqnarray}
&& R_{0}\left[ -6a^{2}-4a(b-1)-2(b-1)b+R_{0}\right]
-2C_{2}r^{2}\left( \frac{R_{0}}{r^{2}}\right) ^{n}\left[
4n^{2}(2a+b)-2(2b+3)n(2a+b)+6(a+b)^{2}-R_{0}\right]
\nonumber\\
&&-g_{0}(q-1)qr^{3}R_{0}L_{m}^{q-1}(r)L_{m}^{\prime }(r)=0,
\end{eqnarray}%
\end{widetext}
with the general solution given by
\begin{widetext}
\be
L_{m}(r)=2^{-1/q}
\left\{ \frac{2C_{2}Br^{2}-nR_{0}r^{2n}\left[
-6a^{2}-4a(b-1)-2(b-1)b-2c_{2}g_{0}(q-1)qr^{2}+R_{0}\right] }{%
g_{0}n(q-1)R_{0}r^{2(n+1)}}\right\} ^{1/q},
\ee
\end{widetext}
where $c_2$ is an arbitrary integration constant, and we have denoted
\be
B=R_{0}^{n}\left[ 4n^{2}(2a+b)-2(2b+3)n(2a+b)+6(a+b)^{2}-R_{0}\right] .
\ee

The gravitational Lagrangian of this model takes the form
\begin{equation}
f\left(R,L_m\right)=\frac{R}{2}+\frac{2}{1+2b+4c}C_2R^{(1+2b+4c)/2}+g_0L_m^q.
\end{equation}
Imposing $a=c=\frac{1}{4}(3-2b)$, $C_2 = 1/\Lambda$, we again obtain the
model $f\left(R,L_m\right) = \frac{1}{2}\left(R + R^2/\Lambda^2\right)+g_0L_m^q$, whose $f(R)$ limit, corresponding to $q=1$ and $g_0=1$, has been extensively studied. Therefore, in general, it is clear that in $f\left(R,L_m\right)$ gravity theories two \emph{different} families of
self-consistent Kasner-type string solutions exist.

\subsection{Quasi-Kasner solutions of the field equations} \label{SectVID} 

Finally, we consider the case in which both Kasner conditions, given by
Eqs.~(\ref{Kasner2}) hold \textit{approximately}, so that $a+b+c\approx 1$,
and $a^2+b^2+c^2\approx 1$, but $R_0=a^2+b^2+c^2+(a+b+c)(a+b+c-2)\neq 0$,
implying that the Ricci scalar $R$ is again nonzero inside the string. For the function $g\left(L_m\right)$ we adopt again a power law form, with $g\left(L_m\right)=g_0L_m^q(r)$. 
Under these assumptions the energy-conservation equation~(\ref{cons2})
becomes
\begin{widetext}
\bea
2C_{2}r^{2}\left( \frac{R_{0}}{r^{2}}\right) ^{n}\left[
R_{0}-2(n-1)(3b(c-1)-c+2n)\right]
+R_{0}^{2}-g_{0}(q-1)qr^{3}R_{0}L_{m}(r)^{q-1}L_{m}^{\prime }(r)=0,
\eea
\end{widetext}
with the general solution for $L_m$ given by
\begin{widetext}
\be\label{96}
L_{m}(r)=2^{-1/q}\left\{ \frac{2C_{2}r^{2}R_{0}^{n}\left[
2(n-1)(3b(c-1)-c+2n)-R_{0}\right] -nR_{0}r^{2n}\left[
R_{0}-2c_{3}g_{0}(q-1)qr^{2}\right] }{g_{0}n(q-1)R_{0}r^{2(n+1)}}\right\}
^{1/q},
\ee
where $c_3$ is an arbitrary constant of integration.
\end{widetext}

The thermodynamic parameters of the quasi-Kasner string are obtained as
\begin{widetext}
\be
\rho (r)=\frac{\left\{ -2C_{2}r^{2}(R_{0}/r^{2})^{n}\left[
4(-1+c)(-1+n)n+R_{0}\right] +nR_{0}\left[
-R_{0}+2g_{0}(-1+q)r^{2}L_{m}^{q}(r)\right] \right\} L_{m}^{1-q}(r)}{%
4g_{0}nqr^{2}R_{0}},
\ee
\be
p_r(r)=\frac{L_m^{1-q}(r) \left\{n R_0 \left[R_0-2 g_0 (q-1) r^2 L_m^q(r)\right]-2 C_2 r^2 \left[4 (n-1) n-R_0\right]
   \left(\frac{R_0}{r^2}\right)^n\right\}}{4 g_0 n q r^2 R_0},
\ee
\be
p_{\phi}(r)=\frac{L_m^{1-q}(r) \left\{n R_0 \left[R_0-2 g_0 (q-1) r^2 L_m^q(r)\right]-2 C_2 r^2 \left[8 (c-1) (n-1) n-R_0\right]
   \left(\frac{R_0}{r^2}\right)^n\right\}}{4 g_0 n q r^2 R_0},
\ee
\be
p_z(r)=-\rho (r).
\ee
\end{widetext}

With the use of Eqs.~(\ref{Leff}) and (\ref{Geff}) we obtain for $G_{eff}$
and $\Lambda _{eff}$ the general expressions
\begin{equation}
G_{eff}=\frac{g^{\prime }\left( L_{m}\right) }{1+2C_{2}R_{0}^{n-1}/r^{2(n-1)}%
}=\frac{ g_0qL_{m}^{q-1}(r) }{1+2C_{2}R_{0}^{n-1}/r^{2(n-1)}%
},
\end{equation}%
and
\begin{eqnarray}
&&\hspace{-0.5cm}\Lambda _{eff} =\frac{1}{1+2C_{2}R_{0}^{n-1}/r^{2(n-1)}} \nonumber\\
&&\hspace{-0.5cm}\times \Bigg[\frac{R}{2}+%
\frac{C_{2}}{n}\frac{R_{0}^{n}}{r^{2n}}+  g\left( L_{m}\right) -g^{\prime }\left( L_{m}\right) L_{m}\Bigg]\nonumber\\
&&\hspace{-0.5cm}= \frac{1}{1+2C_{2}R_{0}^{n-1}/r^{2(n-1)}}
\Bigg[\frac{R}{2}+%
\frac{C_{2}}{n}\frac{R_{0}^{n}}{r^{2n}}+ g_0(1-q)L_m^q(r)\Bigg],\nonumber\\
\end{eqnarray}%
respectively.

Again, we may recover the common model with $R$ and $R^2$ terms for an
appropriate choice of the Kasner parameters, but the important point to note
is that, for all cases in which self-consistent Kasner-type string solutions
exist, other more general models are also possible. In particular, there
exist two distinct families of nonvacuum Kasner-type solutions for which the Kasner
conditions are \emph{not} satisfied inside the string core, and specific
relations between the parameters $a$, $b$ and $c$ must be imposed in order
to satisfy the correspondence principle, embodied in Eq.~(\ref{condexp}).

However, an additional family of string-type solutions also exists in which
the Kasner conditions are satisfied \emph{approximately}, but no further
conditions relating $a$, $b$ and $c$ need to be imposed to ensure the
existence of the general relativistic limit. Setting $C_2^{-1} = R_0 =
2\Lambda$ and $n=2$ in Eq.~(\ref{Lm}) is one way to obtain an effective
cosmological constant (to leading order for large $r$) at the level of the
field equations, which is naively consistent with the results of the
previous work in which Linet$-$Tian type solutions (corresponding to the
exterior of a cosmic string embedded in a space-time with $\Lambda \neq 0$
\cite{LinetTian}) were found to exist in $f(R)$ gravity \cite%
{f(R)_strings,f(R)_strings*}.

Thus, it may be hoped that future studies, in
which the local EOM for field-theoretic variables in the matter Lagrangian
are explicitly solved in conjunction with the gravitational field equations,
recover the solutions obtained in \cite{f(R)_strings,f(R)_strings*} as $%
T^{t}_{t}(r) = T^{z}_{z}(r) \rightarrow 0$ for $r \rightarrow \infty$; that is,
as the string energy density tends asymptotically to zero. For small $r$,
one would also hope to be able to relate the parameters in the Kasner metric
to the fundamental field theory parameters (e.g. the symmetry breaking
energy scale, $\eta$, scalar and vector coupling constants, $\lambda$ and $e$, and topological winding number, $|n|$ in the Abelian-Higgs model), directly,
via expressions such as Eq.~(\ref{Lm}).

\subsection{The mass and angular deficit of the string}

In the following we limit our analysis to the quasi-Kasner
case, that is, we assume that $a+b+c\approx 1$, $a^{2}+b^{2}+c^{2}\approx 1$%
, but $R_{0}\neq 0$. The radius $R_s$ of the string can be obtained from the condition $p_r\left(R_s\right)=0$, and it is obtained as a solution of the equation
\bea
&&n R_0 \left[R_0-2 g_0 (q-1) r^2 L_m^q(R_s)\right]\nonumber\\
&&-2 C_2 R_s^2 \left[4 (n-1) n-R_0\right]
   \left(\frac{R_0}{R_s^2}\right)^n=0.
\eea
In the string Lagrangian given by Eq.~(\ref{96}), inside the bracket we have a sum of three terms, the first varying as a function of $r$ as $1/r^{2+2a+b}$, the second as $1/r^2$, and a third constant term. Therefore in the large distance limit the dominant term is $1/r^2$, and we can adopt for the string Lagrangian the approximate expression
\be
L_m(r)\approx \left[\frac{R_0}{n(1-q)}\frac{1}{r^2}\right]^{1/q}.
\ee
For the radius of the string we then obtain the expression
\be
R_s\approx \left\{\frac{2\left[4n(n-1)-R_0\right]}{nR_0^2\left(1+2g_0/n\right)}\right\}^{1/2(n-1)}C_2^{1/2(n-1)}.
\ee

For this value of the radius both the energy density and the tangential
pressure $p_{\phi}$ have nonzero values at the string surface.

The Tolman mass per unit length of the string is 
\begin{widetext}
\begin{eqnarray}
M\left( R_{s}\right) &\approx &\frac{\pi \beta (1-q)}{g_{0}R_{0}^{2}k^{2(1-1/q)}}\left[
\frac{R_{0}}{n(1-q}\right] ^{\frac{1}{q}}(kR_{s})^{3-b-2c-2/q}\Bigg\{ \frac{%
2C_{2}R_{0}^{n}R_{s}^{2(1-n)}\left[ 2(2c-1)(n-1)n-R_{0}\right] }{%
q(b+2c+2n-5)+2}\nonumber\\
&&-\frac{R_{0}^{2}(2g_{0}+n)}{q(b+2c-3)+2}\Bigg\} ,
\end{eqnarray}%
\end{widetext}
where, in order to avoid the singularity in the origin, we have to impose the conditions $b+2(c+1/q)<3$, and $b+2(c+n+1/q)<5$, respectively.
In the same approximation for the $W$ parameter, which determines the angular deficit, we obtain
\begin{widetext}
\begin{eqnarray}
W\left(R_s\right)&=&\frac{\pi \beta (q-1)}{g_{0}R_{0}^{2}k^{2(1-1/q)}}\left[ \frac{R_{0}}{%
n(1-q)}\right] ^{\frac{1}{q}}(kR_{s})^{3-b-2c-2/q}\Bigg\{ \frac{%
2C_{2}R_{0}^{n}R_{s}^{2(1-n)}\left[ 2(4c-5)(n-1)n+R_{0}\right] }{%
q(b+2c+2n-5)+2}\nonumber\\
&&+ \frac{R_{0}^{2}(2g_{0}+n)}{q(b+2c-3)+2}\Bigg\} ,
\end{eqnarray}
\end{widetext}
where we again impose the same conditions in order to avoid the singularity at $r=0$.

By assuming that $C_2=C_0/\Lambda$, where $\Lambda $ is the cosmological constant with values of the order of $\Lambda =10^{-56}\;{\rm cm^{-2}}$, the radius of the string can be estimated as
\be
R_s\approx \left\{2\frac{\left[4(n-1)n-R_0\right]}{nR_0^2\left(1+2g_0/n\right)}\right\}^{1/2(n-1)}\times 10^{28/(n-1)}\; {\rm cm}.
\ee
The string radius thus rapidly decreases with increasing $n$, which determines the $R$-dependence of the gravitational action. 
In addition, very small values of  $4n(n-1)-R_0$ can further reduce the string radius. Alternatively, since $n=1+a+b/2$, very small values of $a$ and $b$ can significantly increase the radius of the string. For $a+b/2=4$, we obtain values of the string radius of order $R_s\approx 10^7\;{\rm cm}$. In the leading order of approximation the mass-radius relation of the quasi-Kasner string is given as
\be
M\left(R_s\right)\propto \left(kR_s\right)^{5-b-2c-2/q-2n},
\ee
or
\be
M\left(R_s\right)\approx 1.35\times 10^{28}\left(kR_s\right)^{5-b-2c-2/q-2n} \;{\rm g/cm}.
\ee
The mass of the string also depends on $q$, giving the matter Lagrangian dependence in the gravitational action. On the other hand, the mass of the string depends on the coefficient $R_0$, which, due to the impositions of the approximate Kasner conditions, may have a very small value. Hence, depending on the adopted values of the model parameters, a very wide range of radii and masses can be generated for cosmic strings in $f\left(R,L_m\right)$ gravity. Similar relations can be obtained for the angular deficit $W$, $W\propto \left(kR_s\right)^{5-b-2c-2/q-2n}$, with the sign of $W$ depending on the adopted numerical values of the string metric parameters. Under the same assumption of an approximate Gaussian matter density profile, the relation $\alpha \approx \left(\mathcal{G}M/c_{light}^2\right){\rm Erf}\left(D/\sigma _0\sqrt {2}\right)$ leads to a great variety of lensing behaviors, due to the strong dependence of the string mass on the model parameters.

\section{Discussions and final remarks} \label{SectVII} 

In this paper we have considered static, cylindrically symmetric
solutions of the gravitational field equations for a string-like
distribution of matter in the $f\left(R,L_m\right)$ modified gravity theory,
in which the gravitational Lagrangian is given by an arbitrary function of
the Ricci scalar and matter Lagrangian. Having determined the gravitational
field equations for a general cylindrically symmetric metric, with arbitrary
dependence of the metric tensor components on the radial distance, $r$, we
have restricted our subsequent analysis to a specific case by adopting the
Kasner metric for the string interior, in which the components of the metric
tensor are proportional to powers of $r$.

Two distinct models have been investigated. In the first, the gravitational
Lagrangian was fixed in the form of an exponential function, $%
f\left(R,L_m\right)=\left(1/\Lambda\right)\exp \left(R/2\Lambda
+L_m/\Lambda\right)$, which obviously reduces to the Einstein$-$Hilbert
Lagrangian in the limit $R/2 +L_m \ll \Lambda$. In the second, $%
f\left(R,L_m\right)$ was self-consistently determined from the field
equations using the string condition, $T_t^t=T_z^z$, or equivalently, $\rho
=-p_z$, where $\rho$ is the mass per unit volume of the string and $p_z$ is
the thermodynamic pressure, together with the correspondence principle, which
requires the Einstein$-$Hilbert limit to exist for appropriate choices of the free model parameters.

In the first case, from the field equations one can obtain a second-order
linear differential equation for the matter Lagrangian, whose solution can
be expressed in terms of the hypergeometric and Laguerre functions. The
knowledge of the matter Lagrangian fully determines the solution of the
field equations, allowing the determination of the energy density and
anisotropic pressures inside the string. Crucially, in this case, the Kasner
parameters are required to satisfy the relation $2a+b>6$, in order to avoid
divergences in the Tolman mass density (i.e. the relativistic mass per unit
length of the string), $M$, and $W$ parameter, which determines the
angular deficit \cite{def}, as $r \rightarrow 0$.

In the second case, by imposing the string conditions for the field
equations in the Kasner metric, we obtain a second-order differential
equation for $f\left(R,L_m\right)$ which allows the determination of the
gravitational Lagrangian density. In this case $f\left(R,L_m\right)$ has an
additive structure, being the sum of a Ricci scalar-dependent function, and
of a matter Lagrangian-dependent function $g\left(L_m(r)\right)$, which is
an arbitrary function of integration. The dependence on the Ricci scalar can
be simplified with the use of the correspondence principle, by requiring
 the standard general relativistic limit of the modified gravity action
(i.e. the Einstein$-$Hilbert action) to exist for appropriate choices of the
arbitrary parameters and functions of the model. In order to determine the
matter Lagrangian dependence one must use the conservation equation of the
theory, which gives a strongly nonlinear differential equation for $%
g\left(L_m(r)\right)$. In order to obtain some explicit solutions of the
field equations we have assumed the simple case in which $%
g\left(L_m(r)\right)=g_0L_m^q(r)$, $q\neq 1$. This choice allows the immediate determination
of the matter Lagrangian from the conservation equation, and of the general solution of the field
equations, for a gravitational model described by a Lagrangian density of
the form $f\left(R,L_m\right)=R/2+\left(C_2/n\right)R^n+g_0L_m^q$, $q\neq 1$, where $%
n=n(a,b,c)$ depends on the parameters $a$, $b$, $c$ of the Kasner metric. In
this case, the function $n(a,b,c)$ may take two different forms,
corresponding to two distinct sets of conditions imposed on the Kasner
parameters, $a$, $b$, $c$, in order to satisfy the correspondence principle.
Thus, depending on the choice of conditions for $a$, $b$ and $c$, two
distinct string-type models can be obtained. An additional family of
solutions, which we refer to as quasi-Kasner solutions, also exists, in
which both sets of Kasner conditions (i.e. those defining both the
Kasner sphere and the Kasner plane \cite{Kasner}) are approximately satisfied,
but for which the Ricci scalar is nonzero inside the string core.

Importantly, for \emph{all} sets of self-consistent Kasner-type string
solutions obtained in our study, the Ricci scalar is of the form $R =
R_0(a,b,c)/r^2$, where $R_0(a,b,c) = 0$ when both sets of Kasner conditions
are satisfied. Thus we recover, in the context of the $f(R,L_m)$ theory of
gravity, the unique $R=0$ vacuum solution obtained in general relativity
\cite{Kasner,exact-sol} and, previously, in $f(R)$ gravity \cite%
{f(R)_strings,f(R)_strings*}. However, the existence of vacuum solutions in
the $f(R)$ theory for which $R = {\rm  const.} \neq 0$, also found in \cite%
{f(R)_strings,f(R)_strings*}, and which correspond to Linet$-$Tian (LT) type
solutions \cite{LinetTian} under the identification of $R = const.$ with the
cosmological constant term in the gravitational field equations, imply that
the Kasner metric fails to capture LT-type solutions in both $f(R)$ and,
more generally, $f(R,L_m)$ gravity.

The physical interpretation of this is that, although cylindrically symmetric
 ``Kasner-type" solutions, for which $R = 4\Lambda = {\rm constant}$ may be identified with the
 cosmological constant, exist in general relativity \cite{Spindel:79}, no such solutions exist
 for the Kasner metric in $f(R) $ and $f(R,L_m)$ gravity. Hence, since the
 (non-Kasner) solutions exhibiting cylindrical symmetry so
far obtained in $f(R)$ gravity correspond to vacuum solutions for which $R =
{\rm constant} \neq 0$ and, since the unique $R=0$ vacuum solution obtained in \cite%
{f(R)_strings,f(R)_strings*} is equally well expressed in terms of the
Kasner metric, as shown in the present study, we may conclude that
non-Kasner metrics are required to capture \emph{all} additional static,
cylindrically symmetric, vacuum solutions (for which $R \neq 0$) in both the $%
f(R) $ and the $f\left(R,L_m\right)$ modified gravity theories.

Observationally, these results are interesting since they provide a much richer
gravitational lensing pattern for strings in $f(R,L_m)$ gravity, as compared with strings in 
general relativity. In particular, both the potential range of string widths and the range of lensing angles 
is far greater than in the standard scenario, and additional distortion of lensing images occurs
due to the space-time curvature, even for vacuum strings. Since these also depend, sensitively, on the values of 
many free model parameters, it may even be hoped that future high-precision observations of anomalous lensing candidates, such 
as CSL-1, may yield data compatible with $f(R,L_m)$ string models.


\section*{Acknowledgments}

We would like to thank the anonymous referees for comments and suggestions that helped us to improve our manuscript. T. H. thanks the Department of Physics of the Sun-Yat Sen
University in Guangzhou, People's Republic of China, for the kind hospitality offered during the
preparation of this work. M. L. thanks Cherry Ch, for making
fantastic coffee.

\appendix

\section{The conservation equation in $f\left(R,L_m\right)$ gravity}
\label{app}

In this appendix we give an explicit proof of the ``non-conservation"
equation~(\ref{noncons1}) in $f\left(R,L_m\right)$ gravity. By taking
the covariant divergence of Eq.~(\ref{field2a}) we first obtain
\begin{eqnarray}  \label{A1}
&&\hspace{-0.5cm}\nabla ^{\mu}f_{R} R_{\mu \nu }+f_R\nabla ^{\mu}R_{\mu \nu}+\left( \nabla
_{\nu}\Box -\Box\nabla _{\nu }\right) f_{R}  \nonumber\\
&&\hspace{-0.5cm}- \frac{1}{2}\Bigg[f_R\nabla ^{\mu}R+f_{L_m}\nabla ^{\mu}L_m -\nabla
^{\mu}f_{L_m}L_m-
f_{L_m}\nabla ^{\mu }L_m \Bigg]g_{\mu \nu } \nonumber\\
&&\hspace{-0.5cm}= \frac{1}{2}\nabla
^{\mu}f_{L_{m}} T_{\mu \nu }+\frac{1}{2}f_{L_m}\nabla ^{\mu}T_{\mu \nu}.
\end{eqnarray}
With the help of Eq.~(\ref{koi}) we then have
\begin{equation}
\left( \nabla _{\nu}\Box -\Box\nabla _{\nu }\right) f_{R}=-R_{\mu \nu}\nabla
^{\mu }f_R,
\end{equation}
while from purely geometric considerations
\begin{eqnarray}
&&f_R\nabla ^{\mu}R_{\mu \nu}-\frac{1}{2}f_R\nabla ^{\mu}Rg_{\mu \nu}\nonumber\\
&&= f_R\nabla ^{\mu }\left(R_{\mu \nu}-\frac{1}{2}g_{\mu \nu}R\right)\equiv 0.
\end{eqnarray}
Therefore, Eq.~(\ref{A1}) reduces to
\begin{equation}
\nabla ^{\mu}T_{\mu \nu}=\frac{1}{f_{L_m}}\nabla
^{\mu}f_{L_m}\left(L_mg_{\mu \nu}-T_{\mu \nu}\right),
\end{equation}
or, equivalently,
\begin{equation}
\nabla ^{\mu}T_{\mu \nu}=\nabla ^{\mu}\ln f_{L_m}\left(L_mg_{\mu \nu}-T_{\mu
\nu}\right).
\end{equation}

\section{Derivation of the Kasner metric in standard general relativity}
\label{app1}

In standard general relativity the vacuum gravitational field equations
satisfy the conditions $R_{\mu}^{\nu}=0$ and $R=0$, respectively. For a
general static cylindrically symmetric metric of the form (\ref{lineelement}%
), the components of the Ricci tensor are given by Eqs.~(\ref{Ricci}). The
equations $R_t^t=R_{\phi}^{\phi}=R_z^z=0$ can be immediately integrated to
give
\begin{equation}  \label{B1}
LKN^{\prime }=C_1,NKL^{\prime }=C_2,NLK^{\prime }=C_3,
\end{equation}
where $C_1$, $C_2$, $C_3$ are arbitrary constants of integration. From the
above equations we obtain
\begin{equation}
\frac{N^{\prime }}{N}=\alpha _1\frac{L^{\prime }}{L},\frac{N^{\prime }}{N}%
=\beta_1\frac{K^{\prime }}{K},\frac{K^{\prime }}{K}=\frac{1}{\gamma _1}\frac{%
L^{\prime }}{L},
\end{equation}
where $\alpha _1=C_1/C_3$, $\beta _1=C_1/C_3$, and $\gamma _1=C_2/C_3$.
The condition $R_r^r=0$ then gives the following equation for $L(r)$,
\begin{equation}
\left(\alpha _1 \gamma _1+\gamma _1+1\right) L(r) L^{\prime \prime
}(r)-\left(\alpha _1 \gamma _1+1\right) L^{\prime 2}=0,
\end{equation}
which has the general solution
\begin{equation}  \label{B4}
\hspace{-0.2cm}L(r)= c_2 \left(\gamma _1 r\right)^{1+\alpha _1+1/\gamma _1}=c_2
\left(\gamma _1 r\right)^{b}=\beta \left(k r\right)^{b-1}r,
\end{equation}
where $c_2$ is an arbitrary integration constant and where we have taken, without any
loss of generality, an arbitrary integration constant to be zero. We have also
denoted $b=1+\alpha _1+1/\gamma _1$. In the last equality in Eq.~(\ref{B4})
we have rescaled the arbitrary integration constants. By denoting $a=\alpha
_1b$ and $c=b/\gamma _1$, we obtain
\begin{equation}
N(r)=N_0\left(\gamma _1 r\right)^{a}, K(r)=K_0\left(\gamma _1 r\right)^{c},
\end{equation}
where $N_0$ and $K_0$ are also arbitrary integration constants. Then, by using
the above expressions for the metric tensor components, Eqs.~(\ref{B1})
impose the restriction $a+b+c=1$ on the constants $a$, $b$, $c$. Finally,
the condition $R=R_0/r^2=0$ requires $R_0=0$ for the vacuum solution and with the use of the
explicit form of $R_0$, given by Eq.~(\ref{ricci}), it follows that in order
for the Kasner metric to represent a static general relativistic vacuum
solution of the gravitational field equations, the coefficients $a$, $b$, $c$
must also satisfy the second Kasner condition $a^2+b^2+c^2=1$.


\begin{thebibliography}{99}
\bibitem{Planckresults} P.~A.~R.~Ade \textit{et al.} [Planck Collaboration],
arXiv:1303.5062 [astro-ph.CO]; 
P.~A.~R.~Ade \textit{et al.} [Planck Collaboration],
arXiv:1303.5076 [astro-ph.CO]; 
P.~A.~R.~Ade \textit{et al.} [Planck Collaboration], arXiv:1405.0874
[astro-ph.GA];
P.~A.~R.~Ade \textit{et al.} [Planck Collaboration], arXiv:1409.2495
[astro-ph.GA].


\bibitem{Bet} M. Betoule et al., arXiv:1401.4064 [astro-ph.CO] (2014).

\bibitem{Str} L. E. Strigari, Physics Reports \textbf{531}, 1 (2013).

\bibitem{1} O. Bertolami, C. G. Bohmer, T. Harko, and F. S. N. Lobo, Phys.
Rev. \textbf{D 75}, 104016 (2007).

\bibitem{2} T. Harko, Phys. Lett. B \textbf{669}, 376 (2008).

\bibitem{fL} T. Harko and F. S. N. Lobo, Eur. Phys. J. C \textbf{70}, 373
(2010).

\bibitem{fLS} T. Harko, F. S. N. Lobo, and O. Minazzoli, Phys. Rev. D \textbf{87}, 047501 (2013).

\bibitem{litfLm} O. Bertolami and A. Martins, Phys. Rev. D \textbf{85},
024012 (2012); Y. Bisabr, Phys. Rev. D \textbf{86}, 044025 (2012); Y.-B.
Wua, Y.-Y. Zhaoa, R.-G. Cai, J.-B. Lu, J.-W. Lu, X.-J. Gao, Phys. Lett. B 
\textbf{717}, 323 (2012); Q. Xu and S.-Y. Tan, Phys. Rev. D \textbf{86},
123526 (2012); J. Wang and K. Liao, Class. Quantum Grav. \textbf{29}, 215016
(2012); D. Puetzfeld and Y. N. Obukhov, Phys. Rev. D \textbf{87}, 044045
(2013); Y. N. Obukhov and D. Puetzfeld, Phys. Rev. D \textbf{87}, 081502
(2013); O. Bertolami, P. Frazao, and J. P{\' a}ramos, JCAP 05 (2013) 029; O.
Minazzoli, Phys. Rev. D \textbf{88}, 027506 (2013); S. Thakur and A. A. Sen,
Phys. Rev. D \textbf{88}, 044043 (2013); T. Harko and F. S. N. Lobo, Phys.
Rev. D \textbf{86}, 124034 (2012); T. Harko, F. S. N. Lobo, M. K. Mak, and
S. V. Sushkov, Phys. Rev. D \textbf{87}, 067504 (2013); N. Tamanini and T.
S. Koivisto, Phys. Rev. D \textbf{88}, 064052 (2013); Y. Bisabr, Gen.
Relativ. Gravit. \textbf{45}, 1559 (2013); R.-N. Huang, arXiv:1304.5309
[gr-qc] (2013); L. Iorio, Class. Quant. Gravit. \textbf{31}, 085003 (2014);
O. Bertolami and J. P{\' a}ramos, Phys. Rev. D \textbf{89}, 044012 (2014); T.
Harko, Phys. Rev. D {\bf 90}, 044067 (2014).

\bibitem{Gal} T. Harko and F. S. N. Lobo, Galaxies \textbf{2}, 410 (2014).




\bibitem{Ki:76} T. W. B. Kibble, J. Phys. A \textbf{9}, 1387 (1976).

\bibitem{Ki:80} T. W. B. Kibble, Phys. Rep. \textbf{67}, 183 (1980).

\bibitem{Ki:82} T. W .B. Kibble, Acta Phys. Polon. B \textbf{13}, 723 (1982).

\bibitem{topological_defects} A. Vilenkin and E. S. Shellard, \textit{Cosmic
strings and other topological defects}, Cambridge University Press (2000);
M. Hindmarsh and T. Kibble, \textit{Cosmic strings}, Rept. Prog. Phys.
\textbf{58}, 477 (1995).

\bibitem{Pr:84} J. Preskill, Ann. Rev. Nucl. Part. Sci. \textbf{34}, 461
(1984). 

\bibitem{Pr:87} J. Preskill, ``Vortices and Monopoles", Lectures presented
at the 1985 Les Houches Summer School (1985).

\bibitem{Gu:81} A. H. Guth, Phys. Rev. D \textbf{23}, 347 (1981).

\bibitem{ParkerBound} E. N. Parker, Astrophys. J. \textbf{160}, 383 (1970);
M.S. Turner, E.N. Parker, and T.J. Bogdan, Phys. Rev. D \textbf{26}, 1296
(1982); E. N. Parker, Astrophys. J. \textbf{163}, 255 (1971); E. N. Parker,
Astrophys. J. \textbf{163}, 279 (1971).

\bibitem{Ad_etal} F. C. Adams \emph{et al}. Phys. Rev. Lett. \textbf{70}, 15
(1993). 

\bibitem{Ze:75} Ya. B. Zel'dovich, I. Yu. Kobzarev and L. B. Okun, Sov.
Phys. JETP \textbf{40}, 1 (1975).

\bibitem{Ze:78} Ya. B. Zel'dovich and M. Yu. Khlopov, Phys. Lett. \textbf{B 79%
}, 239 (1978).

\bibitem{Ma:91} A. Massarotti, Phys. Rev. D \textbf{43}, 346 (1991).

\bibitem{Na:91} Y. Nambu, H. Ishihara, N. Gouda and N. Sugiyama, Astrophys.
J. Lett. \textbf{375}, L35 (1991).

\bibitem{OneScaleAnalytic} A. Vilenkin, Phys. Rev. D \textbf{24}, 2082
(1981); 
T. W. B. Kibble, Nucl. Phys. B \textbf{252}, 227 (1985); Erratum: Nucl.
Phys. B \textbf{261}, 750 (1985); 
D. P. Bennett, Phys. Rev. D \textbf{33}, 872 (1986); Erratum: Phys. Rev. D 
\textbf{34}, 3932 (1986); 
D. P. Bennett, Phys. Rev. D \textbf{34}, 3592 (1986).

\bibitem{OneScaleNumerical} A. Albrecht and N. Turok, Phys. Rev. Lett.
\textbf{54}, 1868 (1985); 
A. Albrecht, and N. Turok, Phys. Rev. D \textbf{40}, 973 (1989);
D.P Bennett and F.R Bouchet, Phys. Rev. Lett. \textbf{60}, 257 (1988);
D.P Bennett, \textit{High resolution simulations of cosmic string evolution:
numerics and long string evolution}, in Formation and Evolution of Cosmic
Strings, G. W. Gibbons, S. W. Hawking and T. Vachaspati, eds., Cambridge
University Press (1990); B. Allen and E.P.S. Shellard, Phys. Rev. Lett.
\textbf{64}, 119 (1990). 


\bibitem{StringGravRad} A. Vilenkin, Phys. Lett. B \textbf{107}, 47 (1981);
T. Vachaspati and A. Vilenkin, Phys. Rev. D \textbf{31}, 3052 (1985);
D. Garfinkle and T. Vachaspati, Phys. Rev. D \textbf{36}, 2229 (1987).

\bibitem{StringGaugeRad} G. Vincent, N.D. Antunes and M. Hindmarsh, Phys.
Rev. Lett. \textbf{80}, 2277 (1998); 
M. Hindmarsh, S. Stuckey and N. Bevis, Phys. Rev. D \textbf{79}, 123504
(2009). 

\bibitem{NambuGoto} T. Goto, Prog. Theor. Phys. \textbf{46}, 1560 (1971).
Y. Nambu, Nucl. Phys. B \textbf{130}, 505 (1977).

\bibitem{StringTheory} M. B. Green, J. H. Schwarz and E. Witten \textit{%
Superstring Theory: Introduction v1}, Cambridge University Press (1988); J.
Polchinski \textit{String Theory: Introduction to the Bosonic String v1},
Cambridge University Press (2005).

\bibitem{FDStrings} A.-C. Davis and T. W. B. Kibble, Contemporary Physics
\textbf{46}, Issue 5 (2005); 
J. Polchinski, [arXiv:hep-th/0412244] (2006);
M. Sakellariadou, Nucl. Phys. Proc. Suppl. \textbf{192-193}, 68 (2009).

\bibitem{Thesis} M. J. Lake, \textit{Cosmic Necklaces in String Theory and
Field Theory}, Ph.D. Thesis, Queen Mary, University of London, U.K. (2010),
https://qmro.qmul.ac.uk/jspui/handle/123456789/523; M. Lake and  J. Yokoyama, JCAP {\bf 09}, 030 (2012): Erratum, 08 (2013) E01; M. Lake and T. Suyama, Phys. Rev. D {\bf 85}, 083521 (2012); M. Lake and J. Ward, JHEP {\bf 1104}, 048 (2011); M. Lake, S. Thomas, and J. Ward, JCAP {\bf 1001}, 026 (2010).

\bibitem{ScalingFDStrings} M. Sakellariadou, J. Cosm. Astropart. Phys.
\textbf{0504}, 003 (2005);
S.-H. Henry Tye, I. Wasserman and M. Wyman, Phys. Rev. D \textbf{71}, 103508
(2005); Erratum-ibid. D \textbf{71}, 129906 (2005).





\bibitem{f(R)_strings} A. Azadi, D. Momeni and M. Nouri-Zonoz, Phys. Lett.
B \textbf{670}, 210 (2008); 

\bibitem{f(R)_strings*} D. Momeni, H. Gholizade, Int. J. Mod. Phys. D \textbf{18}, 1719 (2009).

\bibitem{f(R)_strings**} M. Sharif and S. Arif, Mod. Phys. Lett. A. {\bf 27}, 1250138 (2012).

\bibitem{f(R)_stringsHD} M. Azreg-A{\" i}nou, EPL \textbf{81}, 60003 (2008).

\bibitem{Teleparallel_strings} J. W. Maluf and A. Goya, Class. Quant. Grav.
\textbf{18}, 5143 (2001); 
L. C. Garcia de Andrade, [arXiv:gr-qc/0102086] (2001).

\bibitem{Braneworld_strings} G. Dvali, Ian I. Kogan, and M. Shifman, Phys.
Rev. D \textbf{62}, 106001 (2000);
A. Lue, Phys. Rev. D \textbf{66}, 043509 (2002).

\bibitem{KK_strings} M. Azreg-A{\" i}nou and G. Cl{\' e}ment, Class.
Quantum Grav. \textbf{13}, 2635 (1996);
C. Furtado, F. Moraes and V. B. Bezerra, Phys. Rev. D \textbf{59}, 107504
(1999). 

\bibitem{Lovelock_strings} M. H. Dehghani and N. Bostani, Phys. Rev. D \textbf{75}, 084013 (2007);
J. Z. Simon, Phys. Rev. \textbf{D 41}, 3720 (1990).

\bibitem{GB_strings} M. Azreg-A{\" i}nou and G. Cl{\' e}ment, Class. Quant.
Grav. \textbf{13}, 2635 (1996);
H.-B. Cheng and Y.-Q. Liu, Chinese Phys. Lett. \textbf{25}, 1160 (2008);
M. E. Rodrigues, M. J. S. Houndjo, D. Momeni and R. Myrzakulov, Canadian
Journal of Physics \textbf{92}, 173 (2014). 

\bibitem{BI_strings} M. H. Dehghania, A. Sheykhic and S. H. Hendid, Phys.
Lett. B \textbf{659}, (2008);
G. W. Gibbons and C. A. R. Herdeiro, Phys. Rev. D \textbf{63}, 064006
(2001); 
R. Ferraro and F. Fiorini, J. Phys.: Conf. Ser. \textbf{314}, 012114 (2011).

\bibitem{Bimetric_strings} D. R. K. Reddy, K. S. Adhao and S. D. Katore,
Astrophys. Space Sci. \textbf{301}, 149 (2006);
D. R. K. Reddy, Astrophys. Space Sci. \textbf{286}, 397 (2003);
P. K. Sahoo, Int. J. Theor. Phys. \textbf{48}, 2022 (2009);
G. S. Khadekar, S. D. Tade, Astrophys. Space Sci. \textbf{310}, 47 (2007);
V. U. M. Rao, T. Vinutha and K. V. S. Sireesha, Astrophys. Space Sci.
\textbf{317}, 79 (2008);
P. K. Sahoo and B. Mishra, Int. J. Pure Appl. Math. \textbf{93}, 275 (2014);
P. K. Sahoo and B. Mishra, African Review of Physics \textbf{8}, 0053
(2013);
S. D. Deo, Int. J. Math. Archive \textbf{2}, 1 (2011).

\bibitem{Misc_strings} D. Momeni, Int. J. Theor. Phys. \textbf{50}, 1493
(2011); 
V. U. M. Rao and T. Vinutha, Astrophys. Space Sci. \textbf{325}, 59 (2010);
D. R. K. Reddy and R. L. Naidu, Int. J. Theor. Phys. \textbf{46}, 2788
(2007);
R. C. Furlong, Phys. Rev. D \textbf{38}, 1701 (1988);
O. R. Dando, Ph.D. Thesis, Durham (1999).
%

\bibitem{ScalarTensor_strings} C. Gundlach and M. E. Ortiz, Phys. Rev.
D \textbf{42}, 2521 (1990); 
D. R. K. Reddy, Astrophys. Space Sci. \textbf{286}, 365 (2003);
V. B. Bezerra and C. N. Ferreira, Phys. Rev. D \textbf{65}, 084030 (2005);
V. B. Bezerra, C. N. Ferreira, J. B. Fonseca-Neto and A. A. R. Sobreira,
Phys. Rev. D \textbf{68}, 124020 (2003); 
D. R. K. Reddy, R. L. Naidu and V. U. M. Rao, Astrophys. Space Sci. \textbf{%
306}, 185 (2006);
D. R. K. Reddy, Astrophys. Space Sci. \textbf{286}, 359 (2003);
M. Em'lia and X. Guimar{\v a}es, Class. Quantum Grav. \textbf{14}, 435
(1997); 
D. R. K. Reddy and R. L. Naidu, Astrophys. Space Sci. \textbf{307}, 395
(2007);
D. R. K. Reddy and R. L. Naidu, Astrophys. Space Sci. \textbf{312}, 99
(2007);
D. R. K. Reddy, Astrophys. Space Sci. \textbf{305}, 139 (2006);
C. N. Ferreira, M. E. X. Guimar{\v a}es, and J. A. Helay{\" e}l-Neto,
Nucl. Phys. B \textbf{581}, 165 (2000);
A. Barros and C. Romero, J. Math. Phys. \textbf{36}, 5800 (1995);
K. S. Adhav, A. S. Nimkar and M. V. Dawande, Astrophys. Space Sci. \textbf{%
310}, 231 (2007);
V. U. M. Rao, T. Vinutha and K. V. S. Sireesha, Astrophys Space Sci. \textbf{%
323}, 401 (2009);
M. Em{\' i}lia and X. Guimar{\v a}es, Class. Quantum Grav. \textbf{14}, 435
(1997); 
V. B. Johri and G. K. Goswami, Aust. J. Phys. \textbf{36}, 235 (1983).

\bibitem{Dilaton_strings} A. A. Tseytlin and C. Vafa, Nucl. Phys. B \textbf{372}, 443 (1992); 
R. Gregory and C. Santos, Phys. Rev. D \textbf{56}, 1194-1203 (1997).

\bibitem{Torsion_strings} W. M. Baker, Class. Quantum Grav. \textbf{7}, 717
(1990); 
P. S. Letelier, Class. Quantum Grav. \textbf{12}, 471 (1995);
R. A. Puntigam and H. H. Soleng, Class. Quant. Grav. \textbf{14}, 1129
(1997); 
P. S. Letelier, Class. Quant. Grav. \textbf{12}, 2221 (1995);
H. H. Soleng, Gen. Relativ. Gravit. \textbf{24}, 111 (1992);
Y.-S Duan and X. Liu, J. High Energy Phys. \textbf{0402}, 028 (2004);
M. L. Ruggiero and A. Tartaglia, Am. J. Phys. \textbf{71}, 1303 (2003);
L. Dias and F. Moraes, Brazilian J. Phys. \textbf{35}, 3A (2005);
L. C. Garcia de Andrade, Gen. Relativ. Gravit. \textbf{35}, 7 (2003);
C. N. Ferreira, Class. Quant. Grav. \textbf{19}, 741 (2002);
A. M. de M. Carvalho, C. Furtado and F. Moraes, Phys. Rev. D \textbf{62},
067504 (2000);
R. A. Puntigam, in \textit{Gravity, Theoretical Physics and Computers:
Proceedings of the Conference on Theoretical Physics, General Relativity and
Gravitation}, Bistritza, Romania (1996);
M. J. S. Houndjo, D. Momeni and R. Myrzakulov, JMPD \textbf{21}, 1250093
(2012); 
R. A. Puntigam,
PhD. Thesis, K{\"o}ln (1996),
http://www.puntigam.de/roland/pdf/thesis.pdf; T. T. Fujishiro, M. J. Hayashi
and S. Takeshita, Mod. Phys. Lett. A \textbf{08}, 491 (1993).


\bibitem{BECstring} T. Harko and M. J. Lake, arXiv:1410.6899 [gr-qc] (2014).

\bibitem{Kasner} E. Kasner, J. Math. \textbf{43}, 217 (1921);
E. Kasner, Trans. A.M.S. \textbf{27}, 155 (1925);
A. Harvey, Gen. Relativ. Gravit. \textbf{22}, 1433 (1990).

\bibitem{exact-sol} D. Kramer, H. Stephani, E. Herlt and M. MacCallum,
\textit{Exact Solutions of Einstein's Field Equations} (Cambridge Univ.
Press, Cambridge, England 1980).

\bibitem{LinetTian} B. Linet, J. Math. Phys. \textbf{27}, 1817 (1986); Q.
Tian, Phys. Rev. D \textbf{33}, 3549 (1986).

\bibitem{Spindel:79} P. Spindel, Gen. Relativ. Grav. {\bf 10}, 699 (1979).

\bibitem{Anderson} M. R. Anderson, \textit{The Mathematical Theory of Cosmic
Strings: Cosmic Strings in the Wire Approximation}, Institute of Physics
Publishing (2003).

\bibitem{no} H.~B.~Nielsen and P.~Olesen,
Nucl.\ Phys.\ B \textbf{61}, 45 (1973). 

\bibitem{VaVo:06} M. Vasilic and M. Vojinovic, Phys. Rev. D \textbf{73},
124013 (2006). 

\bibitem{LaLi} L. D. Landau and E. M. Lifshitz, {\it The Classical Theory of
Fields} (Butterworth-Heinemann, Oxford, United Kingdom, 1998).

\bibitem{Koi} T. Koivisto, Class. Quant. Grav. \textbf{23}, 4289 (2006).

\bibitem{KibbleH} T. W. B Kibble and M. Hindmarsh, Rep. Progr. Phys. \textbf{%
58}, 477 (1995). 

\bibitem{clv} M. Christensen, A. L. Larsen and Y. Verbin, Phys. Rev. D 
\textbf{60}, 125012 (1999).

\bibitem{def} Y. Verbin, Phys. Rev. D \textbf{59}, 105015 (1999).

\bibitem{Marder2} L. Marder, Proc. Roy. Soc. London A \textbf{252}, 45
(1959). 

\bibitem{Bonnor} W. B. Bonnor, J. Phys. A \textbf{12}, 847 (1979).

\bibitem{FuGa:88} T. Futamase and D. Garfinkle, Phys. Rev. D \textbf{37},
2086 (1988). 

\bibitem{GaLa:89} D. Garfinkle and P. Laguna, Phys. Rev. D \textbf{39}, 1552
(1989).

\bibitem{LaGa:89} P. Laguna and D. Garfinkle, Phys. Rev. D \textbf{40}, 1011
(1989). 

\bibitem{Ra:90} A. K. Raychaudhuri, Phys. Rev. D \textbf{41}, 3041 (1990).

\bibitem{BoCh:90} B. Boisseau, C. Charmousis, and B. Linet, Phys. Rev. 
D \textbf{55}, 616 (1997).

\bibitem{Vil1} A. Vilenkin, Phys. Rev. D \textbf{23}, 852 (1981).

\bibitem{Gott} J. R. Gott, Astrophys. J. \textbf{288}, 422 (1985).

\bibitem{Hiscock} W. A. Hiscock, Phys. Rev. D \textbf{31}, 3288 (1985).

\bibitem{Linet1} B. Linet, Gen. Relativ. Gravit. \textbf{17}, 1109 (1985).

\bibitem{Garfinkle1} D. Garfinkle, Phys. Rev. D \textbf{32}, 1323 (1985).

\bibitem{Bozza} V. Bozza and L. Mancini, Mon. Not. R. Astron. Soc. {\bf 356}, 1. p. 371 (2005).


\bibitem{Saz} M. V. Sazhin,O. S. Khovanskaya, M. Capaccioli, G. Longo, M. Paolillo, G. Covone, N. A. Grogin and E. J. Schreier, Mon. Not. R. Astron. Soc. {\bf 376}, 1739 (2007).
\bibitem{Lens} M. V. Sazhin, O. S. Sazhina,
M. Capaccioli, G. Longo, M. Paolillo, and G. Riccio, The Open Astronomy Journal, {\bf 3}, 200 (2010); J. K. Bloomfield and  D. F. Chernoff, Phys. Rev. D {\bf 89}, 124003 (2014).


%
%



\end{thebibliography}
\end{document}